\documentclass[superscriptaddress,preprintnumbers,amsmath,amssymb,prx,twocolumn]{revtex4-2}
\usepackage[utf8]{inputenc}
\usepackage{enumitem}
\usepackage{graphicx,amsmath,color,multirow}
\def\k{\kappa}

\def\a{\alpha}
\def\w{\omega}
\def\W{\Omega}
\def\DD{\Delta}

\def\bk{{\bf k}}

\def\bq{{\bf q}}

\def\bR{{\bf R}}

\def\bt{\boldsymbol{\tau}}

\def\e{\epsilon}

\def\<{\langle}
\def\>{\rangle}
\def\D{\partial}

\let\hide\iffalse
\let\unhide\fi
\usepackage{braket} 

\usepackage[caption=false]{subfig}

\begin{document}

\title{Anharmonic lattice dynamics via the special displacement method}

\author{Marios Zacharias}
\email{zachariasmarios@gmail.com}
\affiliation{Univ Rennes, INSA Rennes, CNRS, Institut FOTON - UMR 6082, F-35000 Rennes, France}
\author{George Volonakis}
\affiliation{Univ Rennes, ENSCR, INSA Rennes, CNRS, ISCR - UMR 6226, F-35000 Rennes, France}
\author{Feliciano Giustino}
\affiliation{ Oden Institute for Computational Engineering and Sciences, The University of Texas at Austin,
Austin, Texas 78712, USA
}%
\affiliation{Department of Physics, The University of Texas at Austin, Austin, Texas 78712, USA}
\author{Jacky Even}
\email{jacky.even@insa-rennes.fr}
\affiliation{Univ Rennes, INSA Rennes, CNRS, Institut FOTON - UMR 6082, F-35000 Rennes, France}

\date{\today}

\begin{abstract}
On the basis of the self-consistent phonon theory and the special displacement method, we develop an approach for the 
treatment of anharmonicity in solids. We show that this approach enables the efficient calculation of temperature-dependent 
anharmonic phonon dispersions, requiring very few steps to achieve minimization of the system's free energy. 
We demonstrate this methodology in the regime of strongly anharmonic materials which exhibit a multi-well 
potential energy surface, like cubic SrTiO$_3$, CsPbBr$_3$, CsPbI$_3$, CsSnI$_3$, and Zr. Our results 
are in good agreement with experiments and previous first-principles studies relying on stochastic 
nonperturbative and molecular dynamics simulations. We achieve a very robust workflow by using harmonic phonons 
of the polymorphous ground state as the starting point and an iterative mixing scheme of the dynamical matrix.
We also suggest that the phonons of the polymorphous ground state might provide an excellent
starting approximation to explore anharmonicity. Given the simplicity, efficiency, and stability of 
the present treatment to anharmonicity, it is especially suitable for use with any electronic structure code 
and for investigating electron-phonon couplings in strongly anharmonic systems.

\end{abstract}

\maketitle

\section{Introduction}
\begin{figure}[t!]
\includegraphics[width=0.43\textwidth]{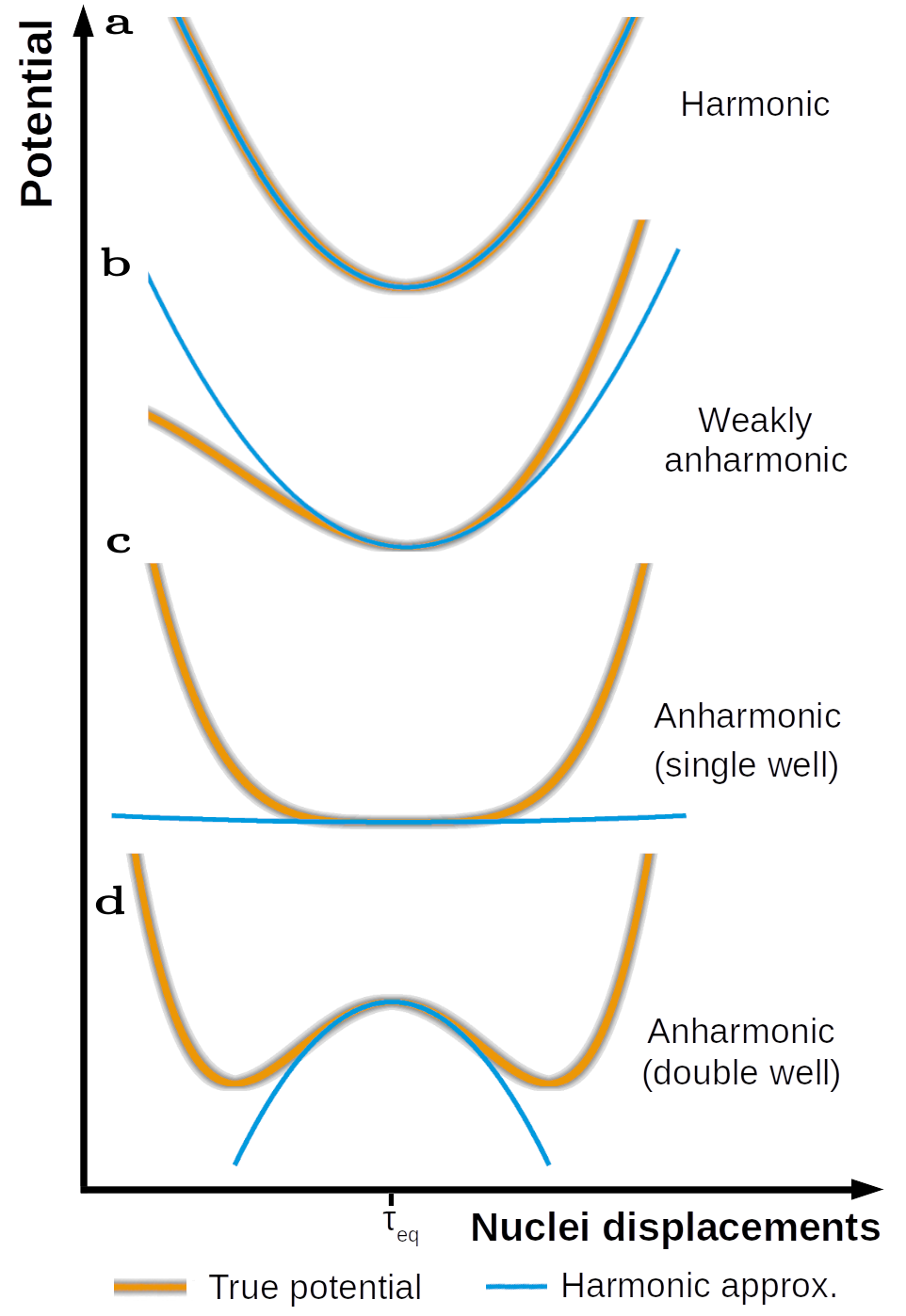}
\caption{ 
Schematic illustration of PES representing various levels of anharmonicity: 
(a) harmonic, (b) weakly anharmonic, (c) anharmonic with a single well, and (d) anharmonic 
with a double well PES. Orange curves represent the true potential and blue curves are harmonic fits around 
static equilibrium positions of the nuclei $\tau_{\rm eq}$. 
\label{fig1} }
\end{figure}

Incorporating anharmonic lattice dynamics in first-principles calculations of solids has been a central topic
of research over the last fifteen years owing to their importance in describing accurately vibrational, transport, 
and optoelectronic properties at finite 
temperatures~\cite{Souvatzis2008,Hellman_2011,Jivtesh2011,Errea_2011,Errea_2013,Zhou_2014,Tadano2014,Rossi_2016,Zacharias2020_FHI}.
Anharmonicity in solids can broadly be classified by inspecting the potential energy surface (PES), representing
the variation of the system's potential energy with respect to the positions of the nuclei (Fig.~\ref{fig1}). 
When variations in the potential energy can be fully or nearly described by 
a parabola [Figs.~\ref{fig1}(a) and (b)], the system is classified as harmonic or weakly anharmonic. 
In these cases one can exploit the standard harmonic approximation 
to efficiently evaluate the vibrational (phonon) spectra by means of density functional perturbation theory (DFPT)~\cite{Baroni2001}
 or the frozen-phonon method (FPM)~\cite{Kunc_Martin}. If, however, the PES is distinctly different from a 
parabola, the system can be classified as anharmonic [Figs.~\ref{fig1}(c) and (d)] and the 
standard harmonic approximation breaks down, leading to instabilities (imaginary frequencies) in the phonon spectrum. 
In this case, one needs to resort to a higher level theory. 

The self-consistent phonon (SCP) theory~\cite{Born1951,Hooton1955,Wethamer_1970} 
forms the basis of current state-of-the-art methodologies employed for the treatment of lattice anharmonicity. 
These methodologies rely on (i) nonperturbative stochastic 
approaches that obtain self-consistency in the phonon spectrum~\cite{Souvatzis2009,Brown2013,Mingo2016,Mingo_2021}, 
or minimize directly the system's free energy~\cite{Errea2014,Patrick2015,Bianco2017,Monacelli2021}, 
(ii) {\it ab initio} molecular dynamics~\cite{Hellman_2013,Tao2014,Bottin2020} (aiMD) for the extraction of an effective matrix 
of interatomic force constants (IFCs) by least-squares fitting, 
and (iii) the explicit computation of higher-order IFCs~\cite{Errea_2011,Tadano_2015,Erba2019,Zhao_2021}.
The success of (i)-(iii) has been demonstrated, among others, 
in calculations of temperature-dependent (anharmonic) phonon dispersions, 
free energies, phase diagrams, lattice conductivities, specific heats, and phonon-phonon scattering 
rates~\cite{Souvatzis2009,Brown2013,Mingo2016,Mingo_2021,Errea2014,Patrick2015,Bianco2017,Hellman_2013,Tao2014,Bottin2020,Tadano_2015,Ribeiro_2018,Erba2019,Zhao_2021}. 

In parallel with the development of efficient computational methods to treat anharmonicity in
solids, the field of electron-phonon interactions from first principles~\cite{Giustino_2017}
is growing tremendously popular~\cite{Ha2021,Yang2022,Engel2022,Jon2022,Wang2022,Spaldin_arxiv}. 
Therefore, interfacing the rather challenging calculations of electron-phonon properties~\cite{Ponce_2016_EPW,Gonze2016,Hyungjun2023} 
with practical calculations of anharmonic phonons is of paramount importance. 
The increasing importance and popularity of these calculations call for the development of
simple and efficient approaches to anharmonicity that can be applied straightforwardly by end-users 
on top of any electronic structure code. 

In this work we focus towards this perspective and develop a nonperturbative first-principles approach to anharmonicity 
that relies on the combination of the SCP theory and the special displacement 
method~\cite{Zacharias_2016,Zacharias_2020}. We demonstrate that special displacements in small 
supercells can be employed to capture very efficiently the anharmonicity in the IFCs via the computation of an effective PES. 
In particular, we show that self-consistency in the phonon spectra can be achieved in very few steps 
using a single thermally distorted configuration at each iteration and taking the polymorphous 
ground state network~\cite{Zhao_Zunger2020} as the starting point. The overall convergence performance is enhanced by 
an iterative mixing scheme. 
Our suggested workflow is very simple as it relies on basic first-principles tools such as the FPM or DFPT. 
We name our approach A-SDM, standing for anharmonicity (A) using 
the special displacement method (SDM). In order to demonstrate the A-SDM, we report anharmonic phonons
of the cubic perovskites SrTiO$_3$, CsPbBr$_3$, CsPbI$_3$, and CsSnI$_3$, tetragonal SrTiO$_3$, 
as well as body-centered cubic (bcc) Zr.
Our results compare well with experiments and previous calculations based on stochastic, aiMD, 
or other variational approaches that rely on truncated Taylor expansions of the PES. 

The paper is organized as follows. In Sec.~\ref{theory} we describe the theoretical framework of lattice dynamics and 
our methodology to anharmonicity, putting together the concepts underpinning SCP and SDM theories. 
In Sec.~\ref{results} we outline the main procedure of the A-SDM and show
its capability in computing temperature-dependent phonon spectra of cubic SrTiO$_3$, CsPbBr$_3$, CsPbI$_3$, CsSnI$_3$, and Zr.
Section~\ref{method} describes improvements related to the self-consistent scheme and reports 
all computational details. Our conclusions and outlook are provided in Sec.~\ref{conclusions}. 









\section{Theory} \label{theory}

Here we describe the theoretical framework underpinning the A-SDM to deal with 
lattice anharmonicity in crystals. Our approach is conceptually similar 
to other SCP methods developed in the past~\cite{Souvatzis2008,Errea_2013,Tadano2014}; the aspect 
introduced here is that special displacements can be employed to accelerate considerably the 
sampling of the nuclei configuration space.

\subsection{Lattice dynamics}

To describe lattice dynamics we resort to supercell periodic boundary conditions and, at first, 
to the standard harmonic approximation. Hence we take the Taylor expansion of the PES 
up to second order in atomic displacements to write:  
\begin{eqnarray}  \label{eq.PES_exp}
 U^{\{\tau\}} &=& U_0 +
         \frac{1}{2}  \sum_{\substack{p \k \a \\ p' \k' \a'}} \frac{\partial^2 U}{\partial \tau_{p \k \a} 
             \partial \tau_{p' \k' \a'}}  \Delta \tau_{p \k \a} \Delta \tau_{p' \k' \a'}. 
\end{eqnarray}
Here $U_0$ is a local extrema of the PES with the atoms at their static equilibrium positions $\tau_{p \k \a}$
where $p$, $\k$, and $\a$ indices represent the unit cell (position vector $\bR_p$),
the atom (position vector $\bt_\k$), and the Cartesian component, respectively. 
The displacements of the atoms away from their static equilibrium positions are denoted as $\Delta \tau_{p \k \a}$. 

To obtain the phonon spectrum one needs to evaluate the dynamical matrix for lattice vibrations obtained as 
the Fourier transform of the IFCs~\cite{FG_Book}: 
\begin{eqnarray}  \label{eq.Dynmat}
    D_{\k \a,\k'\a'}(\bq)  = 
  \sum_{p'}  \frac{C_{0 \k \a, p' \k' \a'}} {\sqrt{M_\k M_{\k'}}}
 e^{i \bq \cdot {\bf R}_{p'}} e^{i \bq \cdot (  {\boldsymbol{\tau}}_{\k'} -{\boldsymbol{\tau}}_{\k} )},
\end{eqnarray}
where $\bq$ is a wavevector of the reciprocal space, the IFCs matrix elements are defined as 
$C_{p \k \a, p' \k' \a'} = \partial^2 U / \partial \tau_{p \k \a}  \partial \tau_{p' \k' \a'}$, 
and $M_\k$ represents the atomic mass. The subscript $0$ is used to indicate that IFCs 
are calculated with respect to a reference unit cell which is invariant upon translation.
For high-symmetry systems, the IFCs entering Eq.~\eqref{eq.Dynmat} should remain invariant under 
the crystal's symmetry operations $\{ {\bf S} | {\bf v}(S) + {\bf R}_m \}$, where ${\bf S}$ represents 
a rotation (proper or improper) and ${\bf v}(S)$ is a fractional translation associated with ${\bf S}$. 
That is, if the position vectors of two atoms are related by~\cite{Maradudin_1968} 
\begin{eqnarray}\label{eq.sym_pos}
\bt_{P K} = {\bf S} \, \bt_{p \k} + {\bf v}(S) + {\bf R}_m,
\end{eqnarray}
then the following relationship should hold for the matrix elements of the IFCs 
\begin{eqnarray}\label{eq.sym_pos2}
C_{P K \a, P' K' \a'} = \sum_{\beta \beta'} S_{\a \beta} S_{\a' \beta' }  C_{p \k \beta, p' \k' \beta'}.	
\end{eqnarray}
A great advantage of using symmetry operations is that  
the computation of the full dynamical matrix can be obtained by applying finite displacements to a few atoms in the unit cell.

For polar semiconductors, the long-range dipole-dipole (dd) interactions, induced by collective ionic displacements,
are treated by employing the standard non-analytical correction in the limit $\bq \rightarrow 0$. This correction 
to the dynamical matrix is given by~\cite{Gonze1997}:
\begin{eqnarray}  \label{eq.Dynmat_nonanal}
 &&   D^{\rm dd}_{\k \a,\k'\a'}(\bq \rightarrow 0)  = \\ 
 &&  \frac{1}{ \sqrt{ M_\k M_{\k'} } } \frac{4 \pi e^2}{ \W} 
    \frac{\sum_\beta q_\beta Z^*_{\k,\beta\alpha} \sum_{\beta'} q_{\beta'} Z^*_{\k',\beta' \a'} }
    {\sum_{\beta \beta'} q_\beta \epsilon^\infty_{\beta \beta'} q_{\beta'} }, \nonumber 
\end{eqnarray}
where $e$ is the electron charge, $\W$ represents the volume of the unit cell, $Z^*_{\k,\beta \a}$ are the 
matrix elements of the Born-effective charge tensor, and $\epsilon^\infty_{\beta \beta'}$ are
the matrix elements of the high-frequency dielectric constant.
This term leads to the splitting between the longitudinal optical (LO) and transverse optical (TO) modes
at the zone-center. At general $\bq$ points, the long-range dipole-dipole effect can be accounted for 
using the linear response or mixed space approaches described in Refs.~[\onlinecite{Gonze1997}] 
and [\onlinecite{Wang2010}], respectively. 

Having obtained the components of the dynamical matrix, the phonon frequencies $\w_{\bq \nu}$ and polarization vectors 
$e_{\k\a,\nu} (\bq )$ corresponding to wavevector $\bq$ and mode $\nu$ are determined by solving the eigenvalue equation:  
\begin{eqnarray}\label{eq.eig_dyn_mat}
\sum_{\k' \a'} D_{\k \a,\k'\a'}(\bq) e_{\k'\a',\nu} (\bq ) 
 = \w^2_{\bq \nu} e_{\k\a,\nu} (\bq). 
\end{eqnarray}
The frequencies $\w_{\bq \nu}$ define the phonon dispersion in the fundamental Brillouin zone of the crystal's 
unit cell. 

\begin{figure}[b!]
\includegraphics[width=0.48\textwidth]{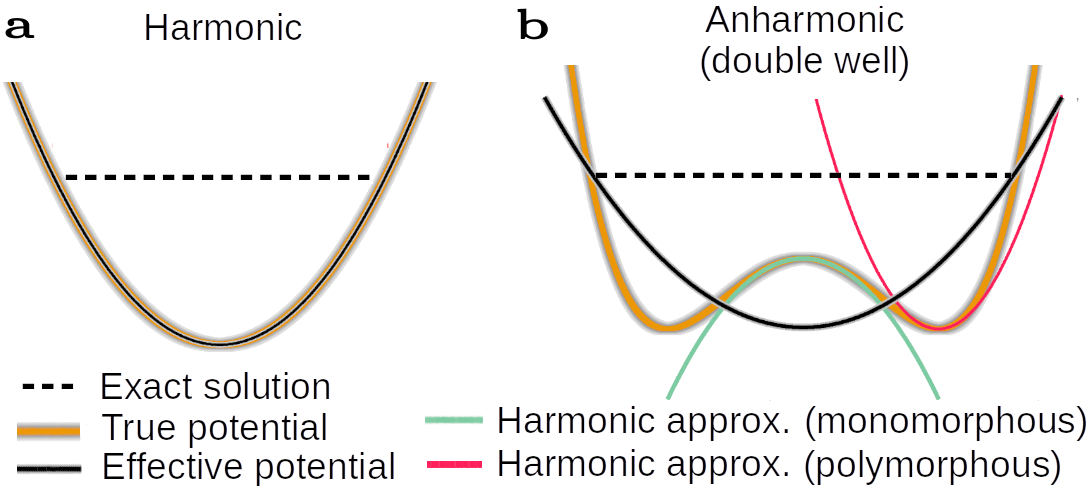}
\caption{ 
Schematic illustration of a harmonic (a) and a double-well anharmonic (b) PES. 
Orange and black curves represent the true and effective potentials. Green and red curves 
are harmonic fits around saddle points of the PES defined by static equilibrium positions of the nuclei.
Black dashed line represents the solution of the nuclear Schr\"odinger equation. \label{fig2}
The local maximum and minima of the PES in (b) correspond to the
monomorphous (high-symmetry) and polymorphous configurations, respectively.}
\end{figure}

\subsection{Treatment of anharmonicity via the special displacement method: A-SDM}

When the displacements are small, the atoms remain near the bottom of the PES well and the harmonic approximation to
lattice dynamics is appropriate for solving the nuclear Schr\"odinger equation [Fig.~\ref{fig2}(a)]. In the case of 
a double well potential, a harmonic approximation at the local maximum is still possible but fails completely to describe 
the free energy of the system [green curve in Fig.~\ref{fig2}(b)]. Instead, one can apply the harmonic 
approximation in one of the two PES wells, corresponding to the system's ground state, and obtain a reasonable 
estimate [red curve in Fig.~\ref{fig2}(b)]. However, there is no direct indication how accurate this estimate is, 
until the actual solution of the nuclear Schr\"odinger equation or the system's free energy are obtained~\cite{Kapil2019}. 
The most popular approach to capture anharmonicity is to find an effective harmonic potential 
whose solution matches the one of the true PES [black line in Fig.~\ref{fig2}(b)].  

To account for anharmonic effects in the lattice dynamics we base our approach on the SCP theory. The merit of this theory 
is that the true potential (e.g. double well) can be replaced by an effective temperature-dependent harmonic
potential, provided this effective potential minimizes the free energy.
The problem involved is to determine, essentially, the temperature-dependent matrix of IFCs 
\begin{eqnarray}  \label{eq.IFC_T}
C_{p \k \a, p' \k' \a'}(T) = \Braket{ \frac{\partial^2 U^{\{\tau\}}} { \partial \tau_{p \k \a}  \partial \tau_{p' \k' \a'}}}_T,
\end{eqnarray}
iteratively until self-consistency is reached. The notation $\braket{.}_T$ represents the ensemble thermal average 
which acts as the trace over the complete set of quantum harmonic oscillators 
weighted by the standard Boltzmann factor at temperature $T$ and normalized by the canonical partition function. 
For completeness, we provide the proof of Eq.~\eqref{eq.IFC_T} in Appendix~\ref{app.proof}.
The thermal average can be expressed as a multivariate Gaussian integral of the following form~\cite{Zacharias_2020}:
\begin{eqnarray}\label{eq_SDM_TA}
     C_{p \k \a, p' \k' \a'}(T)  = 
     \prod_{\bq\nu} \!\int\! \frac{dz_{\bq \nu} }{u_{\bq  \nu} \sqrt{2\pi}  } 
    e^{-\frac{| z_{\bq \nu}|^2}{2 u^2_{\bq  \nu}}} 
   \frac{\partial^2 U^{\{\tau\}}} { \partial \tau_{p \k \a}  \partial \tau_{p' \k' \a'}}. \nonumber  \\
 \end{eqnarray}
Here $z_{\bq \nu}$ are the normal coordinates and $u^2_{\bq  \nu}$ is the mode-resolved
mean-square displacement of the nuclei given by:
\begin{eqnarray}\label{eq_u_qn}
u^2_{{\bq} \nu}(T)  = \frac{\hbar}{2M_0 \omega_{{\bq} \nu}}[2n_{{\bq} \nu}(T) + 1 ],
\end{eqnarray} 
where $M_0$ is the proton mass and $n_{\bq \nu}(T)$ is the Bose-Einstein occupation of the phonon  $\omega_{\bq \nu}$.

At each iteration the thermal average can be evaluated stochastically using Monte 
Carlo approaches~\cite{Errea2014,Zacharias2015,Patrick2015,Mingo_2021}. In these approaches,
displaced nuclei configurations at each temperature are generated by drawing normal coordinates 
from the multivariate normal distribution to evaluate $C_{p \k \a, p' \k' \a'}(T)$ and, hence,  
the dynamical matrix. Minimization of the system's free energy is achieved when self-consistency
with respect to IFCs and thus phonon frequencies and eigenvectors is obtained. This aspect has been demonstrated
and discussed in Ref.~[\onlinecite{Mingo_2021}]. The free energy is defined as $F=U-TS$, where U is the total
energy and $S$ the vibrational entropy, and can be expressed as~\cite{FG_Book,Errea_2013}:
\begin{eqnarray}  \label{eq.FE}
F(T) &=&  \braket{U}_T -  \frac{M_0}{2} \sum_{\bq \nu} \omega^2_{\bq \nu} u^2_{{\bq} \nu}(T) \\
&+& \sum_{ \bq \nu} \bigg[ \frac{\hbar \omega_{\bq\nu}}{2} - k_{\rm B} T\, {\rm ln} [1+ n_{\bq \nu}(T)] \bigg].
\nonumber
\end{eqnarray}
Here, $\braket{U}_T$ is the thermal average of the true PES taken with respect to the eigenstates of 
the effective harmonic Hamiltonian, the term $M_0/2 \sum_{\bq \nu} \omega^2_{\bq \nu} u^2_{{\bq} \nu}(T)$ is the 
vibrational energy of the effective harmonic Hamiltonian, and $k_{\rm B}$ is the Boltzmann constant.

To evaluate Eq.~\eqref{eq.IFC_T} more efficiently we propose to replace the cumbersome stochastic sampling for 
the calculation of $C_{p \k \a, p' \k' \a'}(T)$ with SDM~\cite{Zacharias_2016,Zacharias_2020}. 
For this purpose it suffices to set the nuclei of the system at coordinates defined by 
Zacharias-Giustino (ZG) displacements and calculate
\begin{eqnarray}  \label{eq.ZG_IFC}
C_{p \k \a, p' \k' \a'}(T) \simeq  \frac{\partial^2 U^{\{\tau^{\rm ZG}\}}} { \partial \tau_{p \k \a}  \partial \tau_{p' \k' \a'}}.
\end{eqnarray}
Here the coordinates are defined as $ \tau^{\rm ZG}  =  \tau   +  \Delta \tau^{\rm ZG} $, where $\Delta \tau^{\rm ZG}$ represent 
the ZG displacements. These are given by~\cite{Zacharias_2020}: 
\begin{equation}\label{eq.realdtau_method00}
   \DD\tau^{\rm ZG}_{p \k\alpha} =  \sqrt{\frac{M_0}{N_{p} M_\k}} 2 \, \!\sum_{\bq\nu}\! 
   S_{\bq  \nu} u_{\bq \nu}  \, {\rm Re}
   \Big[ e^{i\bq \cdot {\bf R}_p} e_{\k\alpha,\nu} (\bq ) \Big],
 \end{equation}
where the notation $N_p$ stands for the number of unit cells comprising the supercell. 
The quantities $S_{\bq  \nu}$ represent an optimal choice of signs ($\pm 1$) determined by the \texttt{EPW/ZG} 
module~\cite{Hyungjun2023} so as to ensure that the resulting ZG configuration at temperature $T$ yields 
the best possible approximation to Eq.~\eqref{eq_SDM_TA}.
We note that ZG displacements have been originally designed for the one-shot evaluation of electron-phonon 
effects using supercells. In this work we expand this idea for the efficient treatment of anharmonic dynamics. 
In fact, ZG displacements are perfectly suited for describing an effective potential since one of the premises 
of the SDM theory requires vanishing IFCs between atoms in distant unit cells 
(see Eq.~(45) of Ref.~[\onlinecite{Zacharias_2020}]). 

\begin{figure*}[hbt!]
\includegraphics[width=0.95\textwidth]{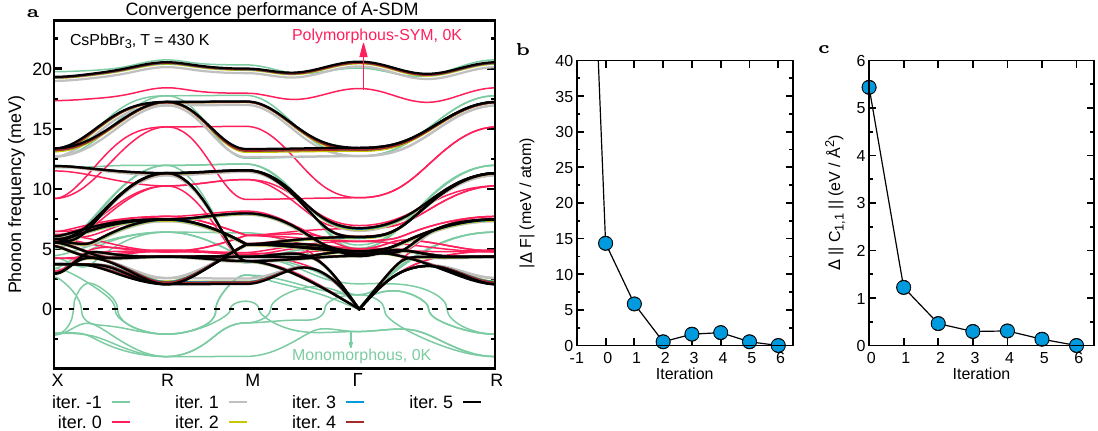}
\caption{(a) Convergence performance of the A-SDM in computing the phonon dispersion of cubic CsPbBr$_3$ at $T=430$~K. 
The calculations refer to 2$\times$2$\times$2 supercells using the experimental lattice constant~\cite{Natarajan1971}
of 5.847~\AA. IFCs at each iteration were evaluated by finite differences 
including corrections due to long-range dipole-dipole interactions. Iteration -1 and 0 correspond
to the phonons calculated for the high-symmetry (monomorphous) and ground state (polymorphous) structures, respectively.  
The dispersion of the polymorphous structure is obtained after applying the crystal's symmetry 
(SYM) operations on the IFCs [Eq.~\eqref{eq.sym_pos2}].
(b) Differential free energy ($\Delta F$) as a function of the number of iterations. 
$\Delta F$ is evaluated as the difference between the free energy [Eq.~\eqref{eq.FE}] 
of the current and the last ZG configuration. (c) Differential Frobenius norm of the leading IFCs 
($\Delta ||C_{1,1}||$) as a function of the number of iterations. $\Delta ||C_{1,1}||$ is evaluated as the difference 
between $||C_{1,1}||$ [Eq.~\eqref{Frob_norm}] of the current and the last IFCs. 
\label{fig3}}
\end{figure*}

The straightforward evaluation of Eq.~\eqref{eq.ZG_IFC} involves performing FPM 
or DFPT on a thermally displaced configuration in the same 
fashion with standard phonon calculations of harmonic systems. However, 
the ``brute force'' computation of $C_{p \k \a, p' \k' \a'}(T)$ by DFPT or FPM becomes prohibitively expensive 
with the supercell size. Instead, one can calculate $C_{p \k \a, p' \k' \a'}(T)$ at each iteration 
much more efficiently by relying only on the Hellman-Feynman forces, $F_{p\k\a}$, acting on the nuclei contained in 
a ZG supercell through the following relationship~\cite{Brown2013,Bianco2017}:
\begin{eqnarray}\label{eq.C_efficient}
C_{p \k \a, p' \k' \a'}(T) &\simeq&  
 \sum_{p'' \k'' \a''} \frac{ \sqrt{M_{\k'} M_{\k''}}}{M_0} \DD\tau^{\rm ZG}_{p'' \k'' \a''}  F_{p \k \a}^{\{\tau^{\rm ZG}\}} \nonumber
 \\ &\times& \sum_{\nu} \frac{e_{\k'\a',\nu}({\bf 0}) e_{\k''\a'',\nu} ({\bf 0}) } {u^2_{{\bf 0} \nu}}. 
\end{eqnarray}
Here all quantities refer to supercell calculations ($\bq={\bf 0}$)
and both $e_{\k\a,\nu} ({\bf 0})$ and $u^2_{{\bf 0} \nu}$ are as obtained in the previous iteration.
The derivation of Eq.~\eqref{eq.C_efficient} proceeds by (i) starting from 
$C_{p \k \a, p' \k' \a'} (T) = \Braket{\partial F_{p \k \a} / \partial \tau_{p' \k' \a'}}_T$, 
 (ii) expressing the derivative with respect to atomic displacements into its normal coordinate 
representation~\cite{Zacharias_2020}, (iii) writing the thermal average as in Eq.~\eqref{eq_SDM_TA}, 
(iv)  making use of integration by parts, and (v) replacing the thermal average 
with its ZG analogue. 

\subsection{Applications of the A-SDM beyond the SCP theory for symmetric phases}
 It should be pointed out that the SDM can also be exploited
for the efficient calculation of the generalization of Eq.~\eqref{eq.IFC_T} for higher-order force constants, 
or in other words, the computation of the Gaussian integral appearing in Eq.~\eqref{eq_SDM_TA} for derivatives of 
the potential of any degree.
The calculation of thermal averages of higher-order force constants should be within reach. 
If successful, this additional information will be useful to investigate second-order phase transitions 
 within Landau's framework~\cite{Bianco2017} in future work.
Furthermore, when nuclei are not enforced to respect crystal symmetries, or occupy 
 general Wyckoff positions (i.e. can be treated as free parameters), it is also desirable to minimize 
the free energy with respect to the internal coordinates. This requires: 
\begin{eqnarray} \label{eq.FE_nuclei}
 \Braket{\frac{\partial U} {\partial \tau_{p \k \a}}}_T = 0.
\end{eqnarray}
The solution of the above equation can be obtained, for example, using the Newton–Raphson method~\cite{Brown2013} 
so that at each iteration we update the equilibrium positions through the relationship: 
\begin{eqnarray}\label{eq_milestone_equilpos}
\tau^{\rm new}_{p \k \a} = \tau_{p \k \a} + C^{-1}_{ p \k \a, p \k \a} \,  F_{p \k \a}^{\{\tau^{\rm ZG}\}}
\end{eqnarray}
where $ C^{-1}_{ p \k \a, p \k \a}$ is the inverse of IFCs matrix evaluated through Eq.~\eqref{eq.ZG_IFC}.
We note that there exist other approaches for solving Eq.~\eqref{eq.FE_nuclei} that rely 
on the conjugate gradient method~\cite{Errea2014}. Also in this case, the SDM can be proven useful. 
In the present manuscript, we mostly study high-symmetry systems characterized by special Wyckoff positions
and therefore Eq.~\eqref{eq.FE_nuclei} is respected by construction so that $\tau^{\rm new}_{p \k \a} = \tau_{p \k \a}$
at each iteration.


\hide
Derivation on how to compute IFCs without performing finite differences: 
\begin{equation}\label{eq.IFC_2}
C_{p \k \a, p' \k' \a'} = \frac{\partial^2 U}{ \partial \tau_{p \k \a}  \partial \tau_{p' \k' \a'}} = 
  \frac{\partial }{ \partial \tau_{p \k \a}} \frac{ \partial U} { \partial \tau_{p' \k' \a'}}
  = \frac{\partial F_{p \k \a} }{ \partial \tau_{p' \k' \a'}}
\end{equation}
where $F_{p \k \a}$ are the Hellman-Feynman forces.

Inegral form of thermal average: 
\begin{eqnarray}\label{eq_SDM_2}
     \hspace{-5pt} \mathcal{O}(T)  &=& \Braket{ \mathcal{O}^{\{\tau\}} }_T \\ & =&
     \prod_{\bq\nu} \!\int\! \frac{dz_{\bq \nu} }{\pi u^2_{\bq  \nu}} 
    e^{-| z_{\bq \nu}|^2/ u^2_{\bq  \nu}} O^{\{\tau\}}. \nonumber  
 \end{eqnarray}

We can also express the partial derivative with respect to atomic displacements as:
\begin{eqnarray}\label{eqa.dtau_partial}
\frac{\D  }{\D \tau_{p \k \a}}  &=& \sum_{\bq \nu} \frac{\D  }{\D z_{\bq \nu }} \frac{\D z_{\bq \nu }}{\D \tau_{p \k a}} 
             \\ &=&
                  N_p^{-1/2} \bigg(\frac{M_\k}{M_0}\bigg)^{1/2} \sum_{\bq \nu} 
                  e^{-i\bq \cdot {\bf R}_p} e^*_{\k\a,\nu}(\bq )\frac{\D  }{\D z_{\bq \nu }} \nonumber.
\end{eqnarray}
We also have:
\begin{eqnarray}\label{eq_partial_TA_identity}
\Braket{ \frac{\D \mathcal{O}^{\{\tau\}} }{\D \tau_{p \k \a}} }_T
=  \frac{\D \braket{\mathcal{O}^{\{\tau\}}}_T }{\D \tau_{p \k \a}} 
\end{eqnarray}

We want to evaluate: 
\begin{eqnarray}\label{eq_partial_TA}
\Braket{ \frac{\D \mathcal{O}^{\{\tau\}} }{\D \tau_{p \k \a}} }_T
=  \prod_{\bq\nu} \!\int\! \frac{dz_{\bq \nu} }{\pi u^2_{\bq  \nu}} 
    e^{-| z_{\bq \nu}|^2/ u^2_{\bq  \nu}} \frac{\D \mathcal{O}^{\{\tau\}} }{\D \tau_{p \k \a}}
\end{eqnarray}
We combine now Eqs.~\eqref{eqa.dtau_partial} and \eqref{eq_partial_TA} to write: 
\begin{eqnarray}\label{eq_partial_TA_2}
\Braket{ \frac{\D \mathcal{O}^{\{\tau\}} }{\D \tau_{p \k \a}} }_T
&=& \bigg(\frac{M_\k}{M_0 N_{\rm p}}\bigg)^{1/2}  
   \sum_{\bq' \nu'} 
   e^{-i\bq' \cdot {\bf R}_p} e^*_{\k\a,\nu'}(\bq' )\nonumber
\\
&& \prod_{\bq\nu} \!\int\! \frac{dz_{\bq \nu} }{\pi u^2_{\bq  \nu}} 
    e^{-| z_{\bq \nu}|^2/ u^2_{\bq  \nu}} \frac{\D \mathcal{O}^{\{\tau\}}  }{\D z_{\bq' \nu' }} 
\end{eqnarray}

Integration by parts: 
\begin{eqnarray}\label{eq_Int_by_Parts}
 v(x) u(x) = \int \bigg[ \frac{\D v(x)}{\D x} u(x) + \frac{\D u(x)}{\D x} v(x) \bigg] dx
\end{eqnarray}

with 
\begin{eqnarray}\label{eq_partial_TA}
 \frac{\D u(x)}{\D x} &=& \frac{\D \mathcal{O}^{\{\tau\}} }{\D z_{\bq' \nu' }} , \, \\
 v(x) &=& 
\frac{e^{-| z_{\bq' \nu'}|^2/ u^2_{\bq'  \nu'}} }{\pi u^2_{\bq'  \nu'}}  
   \prod_{\bq\nu \neq \bq' \nu'} \int \frac{dz_{\bq \nu} }{\pi u^2_{\bq  \nu}} 
    e^{-| z_{\bq \nu}|^2/ u^2_{\bq  \nu}} \nonumber
\end{eqnarray}
and we integrate over $dz_{\bq' \nu'}$. Therefore we have:
\begin{eqnarray}\label{eq_Int_by_Parts_2}
\prod_{\bq\nu} \!\int\! \frac{dz_{\bq \nu} }{\pi u^2_{\bq  \nu}} 
 &&   e^{-| z_{\bq \nu}|^2/ u^2_{\bq  \nu}} \frac{\D \mathcal{O}^{\{\tau\}}  }{\D z_{\bq' \nu' }}
=   \\
&& \frac{e^{-| z_{\bq' \nu'}|^2/ u^2_{\bq'  \nu'}} }{\pi u^2_{\bq'  \nu'}}  
   \prod_{\bq\nu \neq \bq' \nu'} \int \frac{dz_{\bq \nu} }{\pi u^2_{\bq  \nu}} 
    e^{-| z_{\bq \nu}|^2/ u^2_{\bq  \nu}} \mathcal{O}^{\{\tau\}} \bigg|^\infty_{-\infty} \nonumber
\\ 
&& - \prod_{\bq\nu} \!\int\! \frac{dz_{\bq \nu} }{\pi u^2_{\bq  \nu}} 
    e^{-| z_{\bq \nu}|^2/ u^2_{\bq  \nu}} \Big[ \frac{-2 z_{\bq' \nu'}}{u^2_{\bq'  \nu'}}\Big] \mathcal{O}^{\{\tau\}} 
\\
&& = \prod_{\bq\nu} \!\int\! \frac{dz_{\bq \nu} }{\pi u^2_{\bq  \nu}} 
    e^{-| z_{\bq \nu}|^2/ u^2_{\bq  \nu}} \Big[ \frac{2 z_{\bq' \nu'}}{u^2_{\bq'  \nu'}}\Big] \mathcal{O}^{\{\tau\}}
\end{eqnarray}

Therefore Eq.~\eqref{eq_partial_TA_2} reads: 
\begin{eqnarray}\label{eq_partial_TA_2}
\Braket{ \frac{\D \mathcal{O}^{\{\tau\}} }{\D \tau_{p \k \a}} }_T
&=& \bigg(\frac{M_\k}{M_0 N_{\rm p}}\bigg)^{1/2}  
   \sum_{\bq' \nu'} 
   e^{-i\bq' \cdot {\bf R}_p} e^*_{\k\a,\nu'}(\bq' )
\\
&& \prod_{\bq\nu} \!\int\! \frac{dz_{\bq \nu} }{\pi u^2_{\bq  \nu}} 
    e^{-| z_{\bq \nu}|^2/ u^2_{\bq  \nu}} \Big[ \frac{2 z_{\bq' \nu'}}{u^2_{\bq'  \nu'}}\Big] \mathcal{O}^{\{\tau\}}
\nonumber
\end{eqnarray}
We now substitute back the inverse relation:
 \begin{equation}\label{eq.zeta}
   z_{\bq'  \nu'} = \sum_{p' \k' \a'}  \bigg(\frac{M_{\k'}}{M_0 N_{\rm p}}\bigg)^{1/2} 
  e^{-i\bq' \cdot {\bf R}_{p'}} e^*_{\k'\a',\nu'} (\bq' ) \, \DD\tau_{p' \k' \a'} \, .
 \end{equation}
and we have: 
 \begin{eqnarray}\label{eq.zeta_2}
\Braket{ \frac{\D \mathcal{O}^{\{\tau\}} }{\D \tau_{p \k \a}} }_T
&=& 2 \sum_{p' \k' \a'} \frac{ \sqrt{M_\k M_{\k'}}}{M_0 N_{\rm p}}
\\
&\times&   \sum_{\bq' \nu'} 
   e^{- i\bq' \cdot [{\bf R}_p+ {\bf R}_{p'}]}  \frac{e^*_{\k\a,\nu'}(\bq' ) e^*_{\k'\a',\nu'} (\bq' )} {u^2_{\bq'  \nu'}}
\nonumber \\
&& \prod_{\bq\nu} \!\int\! \frac{dz_{\bq \nu} }{\pi u^2_{\bq  \nu}} 
    e^{-| z_{\bq \nu}|^2/ u^2_{\bq  \nu}} \DD\tau_{p' \k' \a'}  \mathcal{O}^{\{\tau\}} \nonumber
 \end{eqnarray}
Thus
 \begin{eqnarray}\label{eq.zeta_2}
\Braket{ \frac{\D \mathcal{O}^{\{\tau\}} }{\D \tau_{p \k \a}} }_T
&=& 2 \sum_{p' \k' \a'} \frac{ \sqrt{M_\k M_{\k'}}}{M_0 N_{\rm p}}
\\
&\times&   \sum_{\bq' \nu'} 
   e^{- i\bq' \cdot [{\bf R}_p+ {\bf R}_{p'}]}  \frac{e^*_{\k\a,\nu'}(\bq' ) e^*_{\k'\a',\nu'} (\bq' )} {u^2_{\bq'  \nu'}}
\nonumber \\
 &&  \Braket{ \DD\tau_{p' \k' \a'}  \mathcal{O}^{\{\tau\}}}_T \nonumber
 \end{eqnarray}
And for the ZG calculation: 
\begin{eqnarray}\label{eq.zeta_2}
  \Braket{ \frac{\D \mathcal{O}^{\{\tau\}} }{\D \tau_{p \k \a}} }_T
  &=& 2 \sum_{p' \k' \a'} \frac{ \sqrt{M_\k M_{\k'}}}{M_0 N_{\rm p}}
  \\
  &\times&   \sum_{\bq' \nu'} 
     e^{- i\bq' \cdot [{\bf R}_p+ {\bf R}_{p'}]}  \frac{e^*_{\k\a,\nu'}(\bq' ) e^*_{\k'\a',\nu'} (\bq' )} {u^2_{\bq'  \nu'}}
  \nonumber \\
   &&  \DD\tau^{\rm ZG}_{p' \k' \a'}  \mathcal{O}^{\{\tau^{\rm ZG}\}} \nonumber
\end{eqnarray}
And take the real part using sets A and B: 
\begin{eqnarray}\label{eq.zeta_2}
  \Braket{ \frac{\D \mathcal{O}^{\{\tau\}} }{\D \tau_{p \k \a}} }_T
  &=&  \sum_{p' \k' \a'} \frac{ \sqrt{M_\k M_{\k'}}}{M_0 N_{\rm p}}
  \\
  &\times& \Bigg[  \sum_{\bq'\in \mathcal{B}, \nu'} 
     2 \Re \bigg[ \frac{e^{- i\bq' \cdot [{\bf R}_p+ {\bf R}_{p'}]}  e^*_{\k\a,\nu'}(\bq' ) e^*_{\k'\a',\nu'} (\bq' )} {u^2_{\bq'  \nu'}} \bigg]
  \nonumber \\
 &+&   \sum_{\bq'\in \mathcal{A}, \nu'} 
     \frac{ \cos\big[ \bq' \cdot ({\bf R}_p+ {\bf R}_{p'} ) \big] e_{\k\a,\nu'}(\bq' ) e_{\k'\a',\nu'} (\bq' )} 
       {u^2_{\bq'  \nu'}}   \Bigg] \nonumber \\
   &&  \DD\tau^{\rm ZG}_{p' \k' \a'}  \mathcal{O}^{\{\tau^{\rm ZG}\}} \nonumber
\end{eqnarray}

 Now applying the above relation on Eq.~\eqref{eq.IFC_2} to compute the IFCs at finite temperature 
\begin{eqnarray}\label{eq.zeta_3}
&& \braket{C_{p' \k' \a', p \k \a}}_T =  \Braket{ \frac{\partial F_{p' \k' \a'} }{ \partial \tau_{p \k \a}}}_T
 =  \sum_{p'' \k'' \a''} \frac{ \sqrt{M_\k M_{\k''}}}{M_0 N_{\rm p}} \nonumber
  \\
  && \times \Bigg[  \sum_{\bq'\in \mathcal{B}, \nu'} 
     4 \Re \bigg[ \frac{e^{- i\bq' \cdot [{\bf R}_p+ {\bf R}_{p''}]}  e^*_{\k\a,\nu'}(\bq' ) e^*_{\k''\a'',\nu'} (\bq' )} {u^2_{\bq'  \nu'}} \bigg]
  \nonumber \\
 && +   \sum_{\bq'\in \mathcal{A}, \nu'} 
     \frac{ \cos\big[ \bq' \cdot ({\bf R}_p+ {\bf R}_{p''} ) \big] e_{\k\a,\nu'}(\bq' ) e_{\k''\a'',\nu'} (\bq' )} 
       {2 u^2_{\bq'  \nu'}}   \Bigg] \nonumber \\
   &&  \DD\tau^{\rm ZG}_{p'' \k'' \a''}  F_{p' \k' \a'}^{\{\tau^{\rm ZG}\}} 
\end{eqnarray}
\unhide

\section{Main procedure and Results}\label{results}

\subsection{Main procedure of the A-SDM} \label{A-SDM}

We stress that the present variant of the SCP theory 
requires only one thermal supercell configuration at each iteration $j$ and avoids 
encountering instabilities in the phonon spectra at any intermediate step; 
this is consistent with the fact that the SCP theory [Eq.~\eqref{eq.IFC_T}] is designed to work
only with positive definite IFCs.
 In the following, we outline the workflow of the proposed technique: 
\begin{enumerate}
\item Obtain the ground state polymorphous configuration of the system in a supercell by tight 
optimization of the nuclear coordinates, keeping the lattice constants fixed. 
\item Compute the initial matrix of IFCs, ${\bf C}_{0}$, 
using the polymorphous configuration by means of 
DFPT or FPM and apply the crystal's symmetries [Eq.~\eqref{eq.sym_pos2}]. 
\item Generate the dynamical matrix [Eq.~\eqref{eq.Dynmat}] and obtain the phonons 
by diagonalization [Eq.~\eqref{eq.eig_dyn_mat}].
\item Generate  $\Delta \tau^{\rm ZG}$ [Eq.~\eqref{eq.realdtau_method00}] for temperature $T$ using the computed phonons and displace the nuclei
from their high symmetry positions in a supercell. This step is performed with the {\tt ZG.x} 
code of {\tt EPW}~\cite{Hyungjun2023}.
\item Compute ${\bf C}_j (T)$ using the ZG configuration 
[either with Eq.~\eqref{eq.ZG_IFC} or Eq.~\eqref{eq.C_efficient}].
\item Apply the crystal's symmetries to ${\bf C}_j(T)$ [Eq.~\eqref{eq.sym_pos2}]. 
\item Apply linear mixing 
${\bf C}_j\,=\,\beta\,{\bf C}_j + (1-\beta)\,{\bf C}_{j-1}$, where $\beta$ is the mixing
factor between 0.2 and 0.7, and compute the phonons as in step 3. 
\item Repeat steps 4-7 until self-consistency in the IFCs [Eq.~\eqref{eq.IFC_T}], and therefore, in 
phonon frequencies and eigenvectors is reached. 
\item If convergence with respect to supercell size is sought, go to step 4 and generate 
a ZG configuration in a larger supercell using the self-consistent phonons obtained in step 8. 
\end{enumerate}

\begin{figure*}[hbt!]
\includegraphics[width=0.75\textwidth]{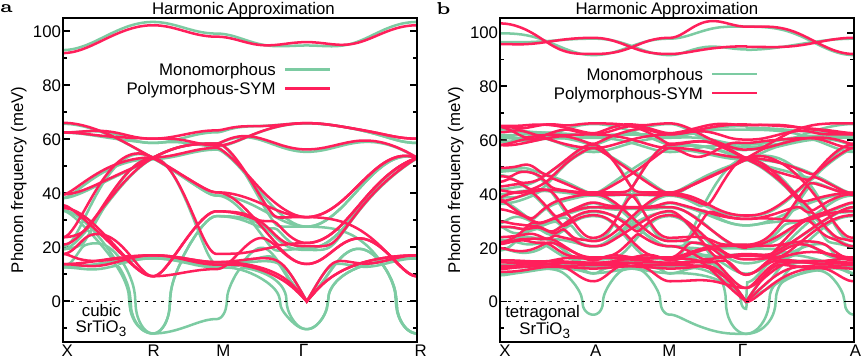}
\caption{Harmonic phonon dispersions of (a) cubic SrTiO$_3$ and (b) tetragonal SrTiO$_3$
calculated using the monomorphous (green) and polymorphous (red) structures.
 Results were obtained using 2$\times$2$\times$2 supercells (40 atoms for the cubic 
and 80 for the tetragonal phase), the PBEsol approximation, and including long-range dipole-dipole interactions.
}
\label{fig4}
\end{figure*}

Let us remark that in the present procedure both phonon frequencies and eigenvectors 
are updated iteratively. However, a complete self-consistent phonon scheme also allows to explore new thermal 
equilibrium positions [Eq.~\eqref{eq_milestone_equilpos}] at each step, which play a 
key role in describing structural phase transitions at critical temperatures~\cite{Bianco2017}. Here, we limit our 
calculations to temperatures away from critical points, and mostly consider systems for which the
nuclei remain at their special Wyckoff positions. 

We note that for materials with a multi-well PES, their ground state structures in step 1 are obtained 
by initially displacing the atoms in a supercell along the soft modes computed for the high-symmetry structure
and then performing a tight relaxation of the atomic coordinates with the lattice constants fixed. 
More details are provided in Sec.~\ref{comp.details}. 
Following the definitions made in Ref.~[\onlinecite{Zhao_Zunger2020}], we refer to the high-symmetry and ground state
structures as the {\it monomorphous} and {\it polymorphous} networks, respectively.

In Fig.~\ref{fig3}(a) we demonstrate the performance of the A-SDM for calculating the phonon dispersion 
of cubic CsPbBr$_3$ at $T=430$~K. 
Iteration -1 refers to the phonons computed for the monomorphous structure, featuring large instabilities. 
Iteration 0 represents the symmetrized phonon dispersion of the polymorphous structure computed by 
performing steps 1-3. We indicate this phonon dispersion as Polymorphous-SYM for the rest of the manuscript. 
The phonon dispersions from iteration 1 to 5 are evaluated 
by repeating steps 4-7 using a mixing parameter $\beta = 0.5$ until self-consistency is achieved. Impressively, our results 
show that the A-SDM exhibits an outstanding efficiency requiring maximum 3-4 configurations for self-consistency 
in the phonon spectra. This can also be seen by inspecting the variation of the differential 
free energy, using Eq.~\eqref{eq.FE},
with respect to the number of iterations, shown in Fig.~\ref{fig3}(b). 
We emphasize that even the first iteration (grey line) provides reasonable results, being very 
close to the converged dispersion. 
We also stress that owing to the linear mixing scheme (step 7), none of the 
iterations suffer from phonon spectra with instabilities. This feature of our technique constitutes an 
advantage over previous approaches~\cite{Brown2013,Mingo_2021}. 
In Sec.~\ref{iter.average}, we illustrate the validity of the iterative mixing scheme. 

As a sanity check, we also calculated the Frobenius norm of the leading components of the 
IFCs matrix as: 
\begin{equation} \label{Frob_norm}
||C_{1,1}|| = \sum_\k ||{\bf C}_{1\k,1\k} ||,
\end{equation}
where we set the unit cell index $p =1$. As shown in Fig.~\ref{fig3}(c), the convergence behavior
of $||C_{1,1}||$ is similar to that of the free energy. 
We propose that computing $||{\bf C}_{1\k,1\k}||$ for small ZG supercells can be advantageous
over relying solely on numerical convergence of the free energy.
This is because the error involved in the evaluation of $\braket{U}_T$, entering the 
expression in Eq.~\eqref{eq.FE}, might slow down convergence for small ZG supercells.






\subsection{SrTiO3}

In this section we demonstrate our methodology by taking as a test case the phonon dispersion of SrTiO$_3$ 
for which extensive theoretical and experimental data exist in the 
literature~\cite{COWLEY_1969,Yamada1969,Stirling_1972,Servoin_1980,Tadano_2015,Zhou_2018,Fumega_2020,Zhao_2021,Mingo_2021,Verdi_2023}.

\begin{figure*}[t!]
\includegraphics[width=0.95\textwidth]{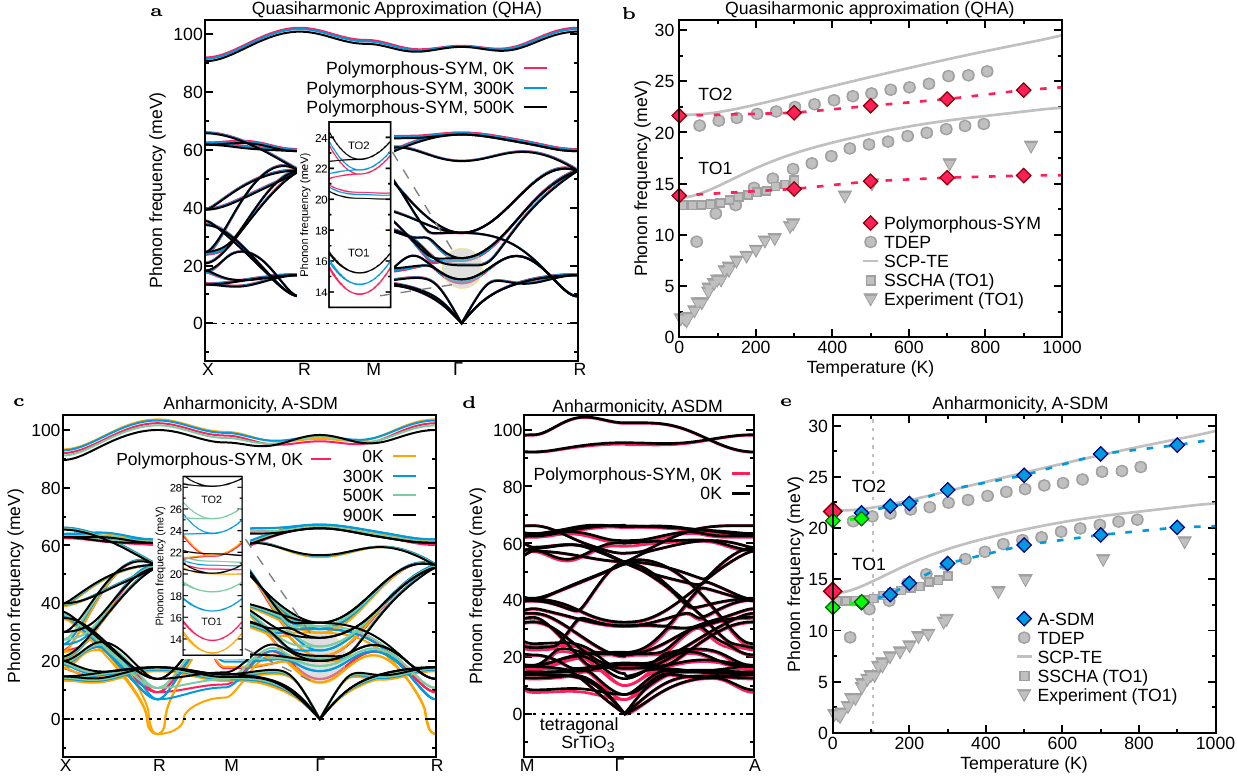}
\caption{
(a) Temperature-dependent phonon dispersion of polymorphous cubic SrTiO$_3$ calculated using 
the quasiharmonic approximation (QHA) for $T = 0$~K (red), $300$~K (blue), and $500$~K (black). 
Thermal lattice expansion was taken into account
using experimental data from Ref.~[\onlinecite{Lingy_1996}]. The phonon dispersions of the polymorphous structures 
were evaluated after applying the crystal symmetries (SYM) to the IFCs. 
(b) Phonon frequency of the zone-center TO modes of cubic SrTiO$_3$ 
as a function of temperature. Red diamonds correspond to our calculations in the QHA
using the cubic polymorphous structure. 
Discs and solid lines represent data from Ref.~[\onlinecite{Tadano_2015}] obtained using the
TDEP and SCP-TE methods, respectively. Squares depict the results of 
Ref.~[\onlinecite{Verdi_2023}] evaluated within the SSCHA.
Experimental data, shown as triangles, are from 
Refs.~[\onlinecite{Yamada1969}] and~[\onlinecite{Servoin_1980}].
(c) Temperature-dependent phonon dispersion of cubic SrTiO$_3$ calculated using the A-SDM 
for $T = 0$~K (orange), $T = 300$~K (blue), $500$~K (green), and $900$~K (black). The symmetrized phonon dispersion 
of polymorphous cubic SrTiO$_3$ at 0~K (red) is shown for comparison. 
(d) Temperature-dependent phonon dispersion of tetragonal SrTiO$_3$ calculated using the A-SDM
for $T = 0$~K (black). The symmetrized phonon dispersion of polymorphous tetragonal SrTiO$_3$ at 0~K (red) is 
shown for comparison.
(e) As in (b) but now blue diamonds represent calculations using the A-SDM. 
Green diamonds are for the TO1 frequency calculated for the tetragonal structure using the A-SDM. 
The vertical dashed line at $T=105$~K indicates the phase transition temperature.
All calculations refer to 2$\times$2$\times$2 supercells of cubic or tetragonal SrTiO$_3$ and 
are performed within the PBEsol approximation. Dispersions in (a), (c), and (d) 
include long-range corrections.
}
\label{fig5}
\end{figure*}



\begin{figure}[htb!]
\includegraphics[width=0.44\textwidth]{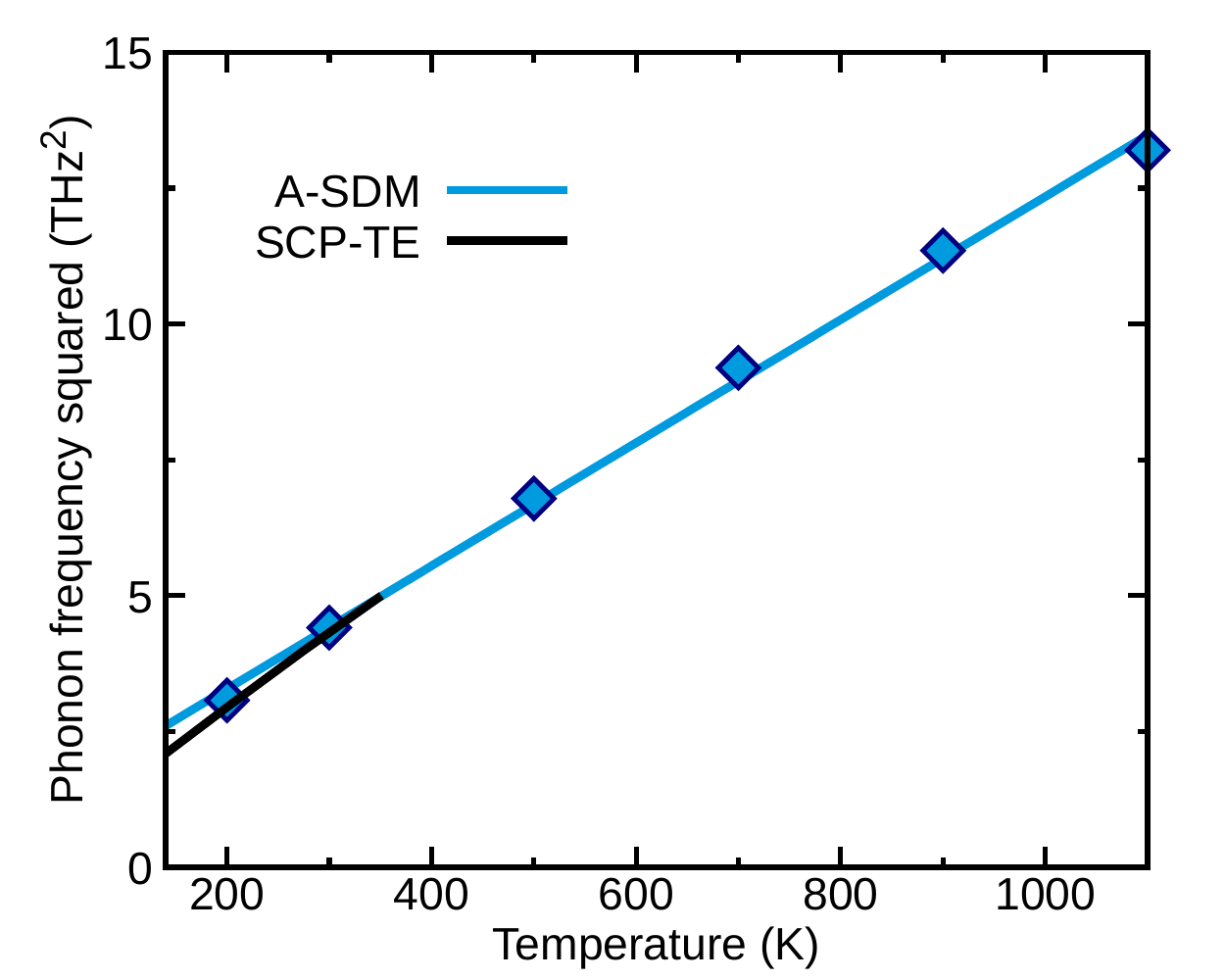}
\caption{
    Temperature dependence of the squared frequency of the AFD soft mode at R-point 
   calculated with the A-SDM (blue diamonds) and SCP-TE (black line) approaches using a 2$\times$2$\times$2 supercell 
   of SrTiO$_3$. SCP-TE data are from Ref.~[\onlinecite{Tadano_2015}]. The linear fit to the A-SDM data is of the 
   form $\omega^2_{\rm R} = 1.014 \, (\text{THz}^2) + 1.133 \, \big(\text{THz}^2 / 100 \text{K} \big)$. }
\label{fig6}
\end{figure}

Figures~\ref{fig4}(a) and (b) show the phonon dispersions of cubic and tetragonal 
SrTiO$_3$, respectively, calculated for the monomorphous and polymorphous structures. Those 
structures correspond to different extrema in the PES as shown schematically in Fig.~\ref{fig2}(b).  
The phonon dispersion of the monomorphous cubic SrTiO$_3$ is dynamically unstable, 
exhibiting imaginary frequencies along the R-M-$\Gamma$ path, including the antiferrodistortive (AFD) and ferroelectric (FE)  
soft modes at R and $\Gamma$ points (green line). Allowing the system to reach its disordered ground state 
[i.e. one of the minima in Fig.~\ref{fig2}(b)], the soft modes are stabilized to real and 
positive frequencies with a significant renormalization of up to 30~meV (red line). 
Similarly, the harmonic phonon dispersion of the monomorphous 
tetragonal SrTiO$_3$ exhibits instabilities, where the soft AFD mode (responsible for the transition 
into the tetragonal phase) is folded onto the zone-center.
The system is stabilized when polymorphism is accounted for, demonstrating that the polymorphous 
networks can be utilized for exploring anharmonicity even in low-symmetry phases.

\begin{figure*}[htt!]
\includegraphics[width=0.97\textwidth]{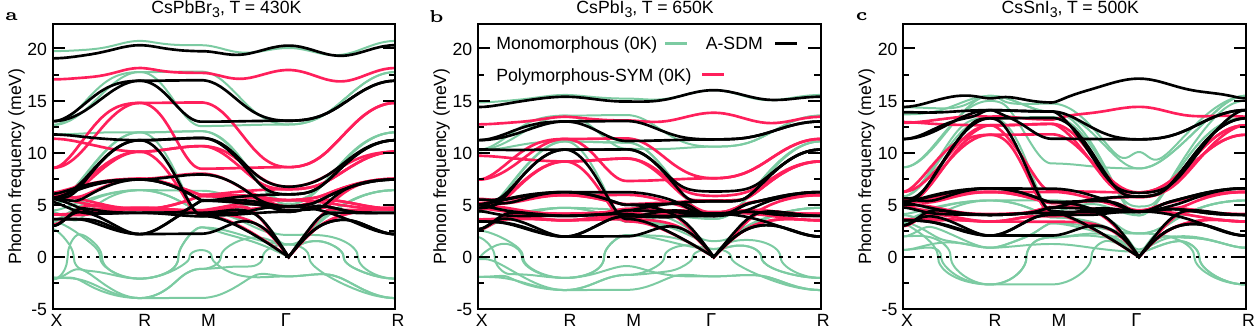}
\caption{ Phonon dispersions of cubic CsPbBr$_3$ at 430~K (a), CsPbI$_3$ at 650~K (b), and CsSnI$_3$ at 500~K (c)
calculated using the A-SDM; all represented by black lines. Green and red lines represent the harmonic phonon dispersions of the 
monomorphous and polymorphous networks at 0~K.  All calculations refer to 2$\times$2$\times$2 supercells 
and include corrections due to long-range dipole-dipole interactions. \label{fig7}}
\end{figure*}

\begin{figure}[b!]
\includegraphics[width=0.45\textwidth]{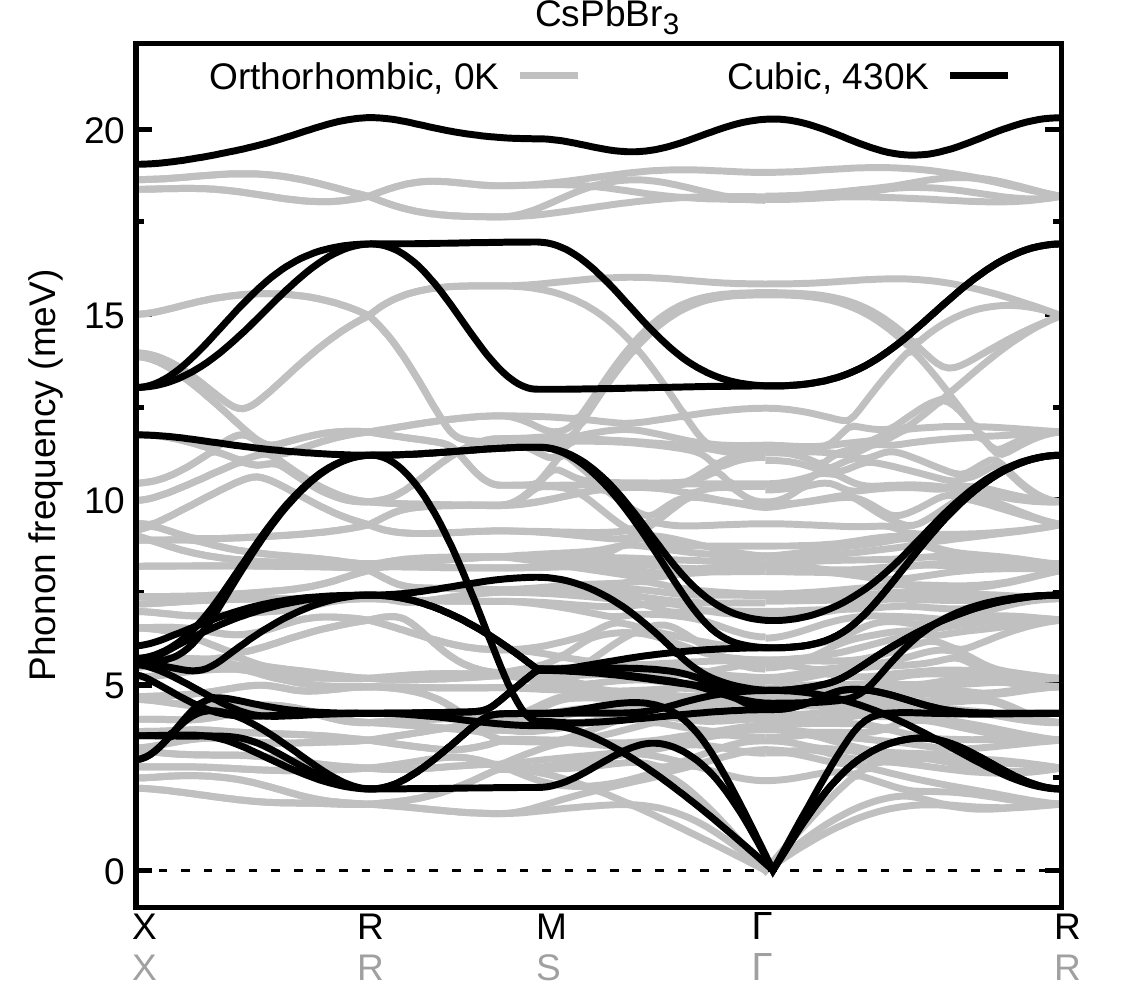}
\caption{ Phonon dispersions of cubic (430~K) and orthorhombic (0~K) CsPbBr$_3$ calculated using the A-SDM 
and FPM, respectively. \label{fig8}}
\end{figure}

As an intermediate step to account for lattice anharmonicity, we compute temperature-dependent phonon dispersions 
of polymorphous cubic SrTiO$_3$ in the quasiharmonic approximation (QHA); i.e. we
repeat phonon calculations for the polymorphous network by varying its volume according to
the measured thermal lattice expansion~\cite{Lingy_1996}.
Figure~\ref{fig5}(a) shows the QHA phonon dispersions at 0~K, 300~K, and 500~K.
The dispersions match closely each other showing that thermal lattice expansion induces small phonon frequency
renormalizations. In the inset, we focus on the effect of thermal lattice expansion on the zone-center TO modes, 
with TO1 representing the FE soft mode. 
In Fig.~\ref{fig5}(b) we present the variation of the TO1 and TO2 soft modes (diamonds) as a function of temperature 
calculated within the QHA. Our values are compared with theoretical data reported 
in Refs.~[\onlinecite{Tadano_2015}] and ~[\onlinecite{Verdi_2023}]. Experimental 
data are from Refs.~[\onlinecite{Yamada1969}] and~[\onlinecite{Servoin_1980}]. 
Theoretical data correspond to calculations performed by combining 
the SCP theory with truncated Taylor expansions of the PES (SCP-TE)~\cite{Wethamer_1970,Tadano_2015} 
(solid lines), using the 
temperature-dependent effective potential (TDEP)~\cite{Hellman_2013} approach (filled discs) that relies on aiMD
simulations, and using the stochastic self-consistent harmonic approximation (SSCHA)~\cite{Monacelli2021} 
(squares) that depends on nonperturbative supercell calculations. {\it It is evident that the QHA combined with phonons of 
the polymorphous structure fails to explain data obtained with higher-level approaches to anharmonicity and experiment.}

Figure~\ref{fig5}(c) shows the temperature-dependent (anharmonic) phonon dispersions of
cubic SrTiO$_3$ at 0~K, 300~K, 500~K, and 900~K calculated within the A-SDM. 
We account for thermal lattice expansion as in the QHA. 
In the inset, we focus on the effect of anharmonicity on the TO1 and TO2 modes.
The soft nature of these modes is evidenced by their frequency sensitivity to
temperature, with TO1 decreasing from 20.07~meV at 900~K to 16.58~meV at 300~K and with TO2 
decreasing from 28.10~meV at 900~K to 23.75~meV at 300~K.
Notably, at $T=0$~K, the AFD mode becomes unstable in the cubic phase for a 2$\times$2$\times$2 supercell
even when anharmonicity is accounted for.
Instead, using the tetragonal structure and including anharmonicity via the A-SDM  
yields stabilized phonon dispersions at $T=0$~K [black line in Fig.~\ref{fig5}(d)] and $T=75$~K (not shown).

To make contact with existing approaches to lattice anharmonicity,
in Fig.~\ref{fig5}(e) we compare our calculations for the temperature dependence of the 
TO1 and TO2 phonon frequencies at the $\Gamma$-point (blue diamonds) with data reported in 
Refs.~[\onlinecite{Tadano_2015}] and ~[\onlinecite{Verdi_2023}].
Our A-SDM values in the temperature range 0--1000~K are consistent with SCP-TE, aiMD, 
and SSCHA data, validating the treatment of lattice anharmonicity within our methodology 
which includes nonperturbatively all anharmonic even-order contributions to the 
phonon self-energy~\cite{Lazzeri2003}. In particular, our calculations for the TO2 phonon frequency as a function 
of temperature are in excellent agreement with the SCP-TE technique. In the latter approach, the phonon frequency 
shifts (i.e. the real part of the phonon self-energy) are evaluated through the loop diagram 
(four-phonon scattering) only, neglecting contributions arising from the bubble (three-phonon scattering) or 
higher-order diagrams. The bubble diagram is presumed to reduce the frequency renormalization of 
the TO1 mode~\cite{Tadano_2015}, and is not captured in the SCP theory, and hence in the A-SDM, 
by construction. Upgrading A-SDM and accounting for the dynamic bubble self-energy requires the explicit evaluation of 
third-order force constants~\cite{Paulatto2015}, but it is beyond the scope of this manuscript.
Furthermore, it is evident that our A-SDM theory captures anharmonicity in the phonon polarization vectors through off-diagonal 
self-energy components since crossing between the TO1 and TO2 frequencies is avoided~\cite{Tadano_2015}.  
Small discrepancies between A-SDM and TDEP phonon frequencies of the order of 1~meV
are attributed to the following possible reasons: (i) the TDEP calculations of 
Ref.~[\onlinecite{Tadano_2015}] neglect thermal expansion of the crystal, (ii) TDEP accounts for the relaxation 
of internal coordinates along the aiMD trajectories which affects the IFCs, (iii)
aiMD simulations generate a non-symmetric distribution of nuclei displacements and
hence odd-order terms in the PES are accounted for, and
(iv) quantum nuclear effects are absent in the TDEP approach as it relies on aiMD.  

To directly compare the A-SDM with the SSCHA data of Ref.~[\onlinecite{Verdi_2023}],
we evaluate the TO1 anharmonic frequency using the tetragonal structure at $T=0$~K and $75$~K, 
as shown by the green diamonds in Fig.~\ref{fig5}(e). This practically demonstrates the agreement of the 
two approaches for two different structural phases. Deviations between the 
two data sets can be explained by the fact that calculations of Ref.~[\onlinecite{Verdi_2023}] are for $4\times4\times4$ 
supercells and include corrections from the, so called, static bubble diagram~\cite{Monacelli2021}.

Figure~\ref{fig5}(e) also shows that the TO1 and TO2 phonon frequencies obtained for the 
polymorphous cubic structure (red diamonds) are in excellent agreement with those calculated 
using the SCP-TE, A-SDM, and SSCHA approaches at 0~K. This result 
together with the phonon dispersions shown in Figs.~\ref{fig5}(c) and (d) demonstrate 
that the polymorphous networks constitute a reasonable 
approximation for dealing with phonon anharmonicity in SrTiO$_3$ at low temperatures.



The temperature dependence of the TO1 phonon frequency calculated with various methods 
within the PBEsol approximation fails to interpret experimental data from 
Refs.~[\onlinecite{Yamada1969}] and~[\onlinecite{Servoin_1980}], as shown in Fig.~\ref{fig5}(e). 
It should be noted that the drop in the frequency observed in the TDEP data is a result of the method's deficiency
in describing quantum nuclear effects, and any improvement with respect to experimental results is merely coincidental.
In reality, the TO1 phonon frequency undergoes a complex frequency shift around the phase transition 
temperature ($\sim 105$~K). This behavior can be described by evaluating 
second derivatives of the SCP free energy~\cite{Bianco2017} and accounting for thermal 
relaxation of the internal nuclei coordinates in a quantum mechanical fashion~\cite{Brown2013}.
However, the key factor that alleviates discrepancy between theory 
and experiment is the use of higher accuracy exchange-correlation functionals~\cite{Wahl_2008}.  
In fact, using the random-phase approximation for the treatment of
 electron correlation effects yields TO1 frequencies in excellent quantitative agreement with
experiment~\cite{Verdi_2023}.

In Fig.~\ref{fig6}, we also report the variation of the AFD soft mode frequency at the 
R-point with temperature, yielding excellent agreement with SCP-TE data obtained for a 2$\times$2$\times$2 
supercell of cubic SrTiO$_3$. 
Employing the tetragonal structure for temperatures below the phase transition, the AFD mode 
at $\Gamma$ splits into a double degenerate $E_g$ mode and an $A_{1g}$ mode. 
Using the experimental lattice constants~\cite{Okazaki1973}, we found the frequency
of the $E_g$/$A_{1g}$ mode to be 3.61/10.71~meV at $T=0$~K and 5.51/10.48~meV at $T=75$~K 
in agreement within 1~meV with those reported in Ref.~[\onlinecite{Verdi_2023}].
These results further validate our A-SDM approach. We remark that 
the frequency of the AFD mode is expected to reduce upon increasing the supercell size~\cite{Tadano_2015,Mingo_2021} 
or changing the exchange-correlation functional~\cite{Verdi_2023}.

\subsection{Metal halide perovskites: CsPbBr$_3$, CsPbI$_3$, and CsSnI$_3$} \label{disord_phonons}
 
In Figs.~\ref{fig7}(a)-(c), we present the phonon dispersions (black lines) of 
cubic CsPbBr$_3$, CsPbI$_3$, and CsSnI$_3$ calculated using the A-SDM for 2$\times$2$\times$2 supercells. 
We note that the A-SDM performance for CsPbI$_3$ and CsSnI$_3$ exhibits a similar convergence rate 
with the one illustrated for CsPbBr$_3$ (Fig.~\ref{fig3}). We choose temperatures for which the  
systems remain thermodynamically stable at their cubic phases~\cite{Hirotsu1974,Trots2008,Yamada1991}.
For comparison purposes we also include the phonons of the monomorphous (green lines) 
and polymorphous (red lines) structures. 
Our results show that IFCs of the monomorphous structures yield
dynamically unstable phonons represented by the negative frequencies in the phonon dispersions (green lines). 
We point out that the high energy modes computed with the monomorphous structures of CsPbI3 and CsPbBr3 
compare well with those obtained from the A-SDM. Polymorphism in metal halide perovskites causes a narrowing 
of the phonon dispersion where the majority of the high energy phonons decrease in energy, leading to enhanced phonon bunching. 
Interestingly, polymorphism induces the hardening of the low frequency soft modes represented by a nearly 
flat band along the entire X-R-M path. Instead, including anharmonicity via the A-SDM, we capture the soft mode behavior 
along the R-M path (black lines) which is consistent with previous calculations~\cite{Patrick2015,LaniganAtkins2021}.  
Moreover, narrowing of the phonon density of states is not present in our A-SDM calculations.  
In Table~\ref{table.4} we provide frequencies of the FE and AFD
soft modes at finite temperatures obtained with the A-SDM theory and compare them with previous 
calculations. Our values are in excellent agreement with SCP and aiMD data from 
Refs.~[\onlinecite{Patrick2015}],~[\onlinecite{LaniganAtkins2021}], and [\onlinecite{Ning2022}]
apart from the $\w_{\rm AFD} (T)$ frequency of CsPbBr$_3$ (2.19~meV compared to 0.95~meV). We tentatively ascribe 
this difference to the sensitivity of the soft mode to the choice of numerical settings, 
like the pseudopotential and kinetic energy cutoff and to the fact that, unlike aiMD simulations, 
our scheme does not allow for thermal disorder via the relaxation of nuclei coordinates. 
As explained in Ref.~[\onlinecite{LaniganAtkins2021}], phonon frequencies 
along the R-M path are very sensitive to variations of the interatomic forces acting between Br atoms, and thereby, 
to changes in the internal nuclei coordinates. 

In Fig.~\ref{fig8} we compare the A-SDM phonon dispersion of cubic CsPbBr$_3$ at 430~K (black) with the 
harmonic one evaluated for orthorhombic CsPbBr$_3$ at 0~K (grey). It can be clearly seen that the two dispersions 
have distinct qualitative and quantitative differences, showing that phonons calculated for the 
orthorhombic structures do not provide a good approximation for describing anharmonicity 
in metal halide  perovskites. For example, using the orthorhombic structure to investigate electron-phonon properties, 
like carrier mobilities or relaxation rates, of high-temperature phases would be inaccurate.

\begin{table}[t!] 
 \captionsetup{font=small}
\caption{Temperature-dependent frequencies of the FE soft mode at the $\Gamma$-point ($\w_{\rm FE}$) and 
        AFD soft mode at the R-point ($\w_{\rm AFD}$) of cubic SrTiO$_3$, CsPbBr$_3$, CsPbI$_3$, and CsSnI$_3$. 
        Present calculations refer to the A-SDM theory; temperatures are indicated below each compound. 
        In the square brackets we provide the values obtained for the polymorphous structures at 0~K. 
        Previous works' data for SrTiO$_3$, CsPbBr$_3$, CsPbI$_3$, and CsSnI$_3$ are from 
         Refs.~[\onlinecite{Tadano_2015}],~[\onlinecite{LaniganAtkins2021}], [\onlinecite{Ning2022}], and~[\onlinecite{Patrick2015}]. 
         All data refer to 2$\times$2$\times$2 supercells except those reported in Ref.~[\onlinecite{Ning2022}] which 
         are for 4$\times$4$\times$4 supercells.  }
        \hspace*{-0.2cm}
\centering
 \begin{tabular*}{0.48\textwidth}{l *{4}c }
 \hline\hline \\  [-0.35 cm]      
  & \multicolumn{2}{c}{Present work} & \multicolumn{2}{c}{Previous works} \\
              & \,\, $\w_{\rm FE} (T)$ \,\, & \,\, $\w_{\rm AFD} (T)$ \,\, & \, $\w_{\rm FE} (T)$ \, & \, $\w_{\rm AFD} (T)$ \,  \\ [0.05 cm]  
            &  (meV)  & (meV)    &  (meV)  & (mev)  \\ [0.05 cm]  \hline \hline \\ [-0.62 cm]\\
 SrTiO$_3$ & \multirow{2}{*}{17.32 [13.86]}    &  \multirow{2}{*}{8.65 [9.32]}   & \multirow{2}{*}{17.86}  &  \multirow{2}{*}{8.56} \\
  (300~K)   &  \\ [0.01 cm] \hline 
 CsPbBr$_3$ &  \multirow{2}{*}{4.34 [4.82]}   & \multirow{2}{*}{2.19 [4.55]}  & \multirow{2}{*}{4.10} &  \multirow{2}{*}{0.95} \\ 
  (430~K) \\ [0.05 cm] \hline
 CsPbI$_3$  & \multirow{2}{*}{3.93 [4.17]}     &  \multirow{2}{*}{1.96 [3.52]}    & \multirow{2}{*}{3.97} & \multirow{2}{*}{2.02} \\ 
  (650~K) \\ [0.05 cm] \hline
 CsSnI$_3$  & \multirow{2}{*}{4.32 [4.44]}      &   \multirow{2}{*}{2.05 [3.40]}   & \multirow{2}{*}{4.26}  & \multirow{2}{*}{2.30}  \\  
  (500~K) \\ [0.05 cm] \hline \hline
 \end{tabular*}
\label{table.4}
\end{table}

\begin{table}[t!] 
 \captionsetup{font=small}
\caption{Average high-frequency dielectric constant ($\e^\infty$), and average atomic Born effective charges ($Z^*_{\k}$)
        of monomorphous (M) and polymorphous (P) SrTiO$_3$, CsPbBr$_3$, CsPbI$_3$, and CsSnI$_3$. 
        Calculations were performed using 2$\times$2$\times$2 supercells.  }
        \hspace*{-0.2cm}
\centering
 \begin{tabular*}{0.45\textwidth}{l *{3}c }
 \hline\hline \\  [-0.35 cm]  
               & \,\,\,\,\,\,\,\,\, $\e^\infty$ \,\,\,\,\,\,\,\,\, & $Z^*_{\k}$   \\ [0.05 cm]  
               &   (e)                 &    \\ [0.05 cm]  \hline \hline \\ [-0.62 cm]\\
 M-SrTiO$_3$   &      6.31       & Sr: 2.56, Ti: 7.32, \,O: -3.29   \\ [0.01 cm] 
 P-SrTiO$_3$   &      6.12       & Sr: 2.55, Ti: 7.00, \,O: -3.18 \\ [0.01 cm] \hline 
 M-CsPbBr$_3$  &      5.31       & Cs: 1.37, Pb: 4.33, Br: -1.90  \\ [0.05 cm]
 P-CsPbBr$_3$  &      4.61       & Cs: 1.31, Pb: 4.06, Br: -1.79   \\ [0.05 cm] \hline
 M-CsPbI$_3$   &      7.12       & Cs: 1.44, Pb: 5.00, \,I: -2.15 \\ [0.05 cm]
 P-CsPbI$_3$   &      5.70       & Cs: 1.33, Pb: 4.42, \,I: -1.92 \\ [0.05 cm] \hline
 M-CsSnI$_3$   &      340.84     & Cs: 1.15, Sn: 4.48, \,I: -1.88 \\ [0.05 cm]
 P-CsSnI$_3$   &      9.91       & Cs: 1.36, Sn: 5.31, \,I: -2.19 \\  [0.05 cm] \hline \hline
 \end{tabular*}

\label{table.3}
\end{table}

Table~\ref{table.3} reports the average high-frequency dielectric constants ($\e^\infty$) 
and atomic Born effective charges ($Z^*_{\k}$) calculated for the
monomorphous and polymorphous 2$\times$2$\times$2 supercells of all cubic compounds using DFPT~\cite{baroni2010}. 
The averages were performed over the diagonal elements of the dielectric constant and effective charge tensors.
It is evident that the cubic polymorphs of SrTiO$_3$, CsPbBr$_3$, and CsPbI$_3$ yield a decrease in the 
absolute values of $\e^\infty$ and $Z^*_{\k}$, with the larger renormalizations found for CsPbI$_3$ which
features the higher degree of polymorphism~\cite{Zacharias2023}. 
It is worth noting that the monomorphous CsSnI$_3$ exhibits an unphysically large high-frequency dielectric constant
due to its metallic-like behavior in density functional theory calculations. 
This leads to a negligible LO-TO splitting at the $\Gamma$-point as shown by the green dispersion 
in Fig.~\ref{fig7}(c). On the contrary, accounting for the polymorphous cubic CsSnI$_3$ yields a 
band gap opening and, thus, a reasonable value for the high-frequency dielectric constant of $\e^\infty=9.91$. 
This value leads to an LO-TO splitting similar to the other metal halide compounds. 

\subsection{Zr}

\begin{figure}[t!]
\includegraphics[width=0.46\textwidth]{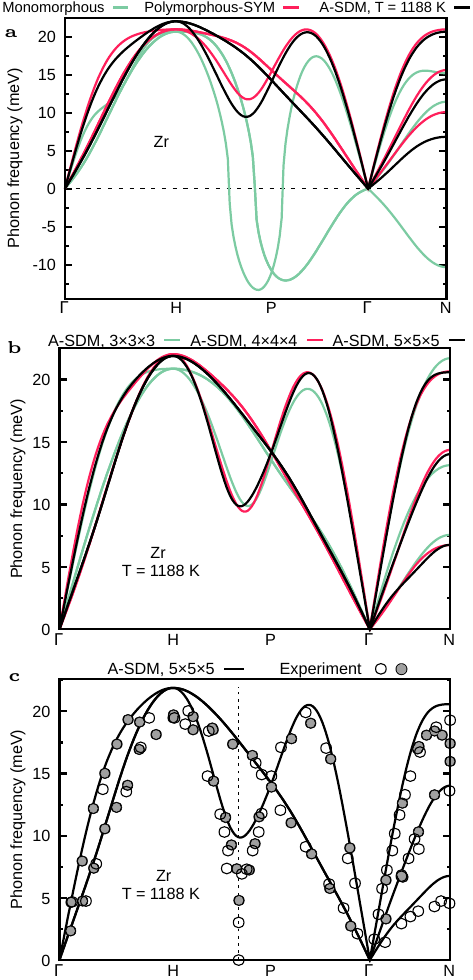}
\caption{(a) Phonon dispersion (black) of bcc Zr at 1188~K calculated using the A-SDM. Harmonic phonon dispersions 
of the monomorphous (green) and polymorphous (red) structures at 0~K are shown for comparison.  
All calculations refer to 4$\times$4$\times$4 supercells. 
(b) Convergence of the A-SDM phonon dispersions of bcc Zr at 1188~K as a function of supercell size. 
Green, red, and black lines represent calculations performed using 3$\times$3$\times$3, 
4$\times$4$\times$4, 5$\times$5$\times$5 supercells, respectively.
(c) Comparison of the phonon dispersion of bcc Zr at 1188~K calculated using the A-SDM for a 5$\times$5$\times$5
 supercell (black line) with neutron scattering measurements from Refs.~[\onlinecite{Stassis1978}] (discs) 
and~[\onlinecite{Heiming1991}] (circles). The vertical dashed line cuts the horizontal axis 
at the $\frac{3}{2} (1,1,1)$ point of the reciprocal space.
\label{fig9}}
\end{figure}

In this section, we study bcc Zr as a representative example of a high temperature phase, 
crystallizing above 1135~K~\cite{Stassis1978}. In Fig.~\ref{fig9}(a), we present the phonon dispersion (black) of 
bcc Zr at 1188~K evaluated using the A-SDM for a 4$\times$4$\times$4 supercell. On the same plot we include 
the symmetrized harmonic phonon dispersions obtained for the monomorphous (green) and polymorphous (red)
structures. Unlike the phonon spectrum of the monomorphous structure, dispersions obtained with the A-SDM or the 
polymorphous network do not suffer from instabilities. Interestingly,
the polymorphous structure constitutes a reasonable approximation for the description of the vibrational 
properties of bcc Zr, yielding phonon dispersions close to those calculated
using the A-SDM. 

To demonstrate convergence we performed calculations using supercells of different size, from 
3$\times$3$\times$3 to 5$\times$5$\times$5, as shown in Fig.~\ref{fig9}(b). Increasing the supercell size is important,
not only for the more precise sampling of the effective PES, but also for approaching the limit where the nonperturbative
calculation with the ZG configuration becomes exact~\cite{Zacharias_2016}. It can be readily seen that the phonon 
dispersion of bcc Zr is well-converged for a 4$\times$4$\times$4 supercell. Notably, self-consistency in the A-SDM 
for 3$\times$3$\times$3 and 4$\times$4$\times$4 supercells was achieved using four iterations.
For the A-SDM calculation in a 5$\times$5$\times$5 supercell we used only a {\it single} ZG-configuration starting 
from the self-consistent phonons obtained for the 4$\times$4$\times$4 supercell (step 9 of the main procedure 
in Sec.~\ref{A-SDM}). This component of our approach provides a significant computational advantage over aiMD
which requires starting from scratch for each supercell size. 

Figure~\ref{fig9}(c) compares our calculated phonon dispersion of bcc Zr at 1188~K for a
 5$\times$5$\times$5 supercell (black) with neutron scattering experiments (circles and discs)~\cite{Stassis1978,Heiming1991}. 
Our A-SDM dispersion is in good agreement with experimental data points across the 
entire $\Gamma$-H-P-$\Gamma$-N high-symmetry path. We note that our results also compare well with 
previous theoretical works that rely on stochastic and aiMD approaches~\cite{Souvatzis2008,Hellman_2011}. 
The A-SDM calculations overestimate the phonon frequency of the longitudinal mode 
at $\frac{3}{2} (1,1,1)$ which is responsible for a martensitic transition towards 
the so-called $\omega$ phase~\cite{Heiming1991}. This mode is, in fact, strongly overdamped and its frequency 
reaches down to zero. Experimental studies~\cite{Stassis1978,Heiming1991} suggest contrasting conclusions on 
whether this behavior arises from a superlattice (elastic) reflection or inelastic scattering. Clarifying 
this aspect from first-principles requires updating the A-SDM to account for the internal relaxation of 
nuclei coordinates under external pressure and evaluating second 
derivatives of the free energy~\cite{Brown2013,Bianco2017}. 

\section{ADDITIONAL METHODOLOGICAL CONSIDERATIONS} \label{method}

\subsection{Importance of using an initial polymorphous structure 
and iterative mixing in the A-SDM.} \label{iter.average}

The main procedure for computing temperature-dependent anharmonic phonons via the A-SDM 
has been described in Sec.~\ref{A-SDM}. Here we discuss some numerical issues if steps 1 and 6 are overlooked.  
We also demonstrate the iterative linear mixing scheme used to speed up convergence. 

\begin{figure*}[htb!]
\includegraphics[width=0.99\textwidth]{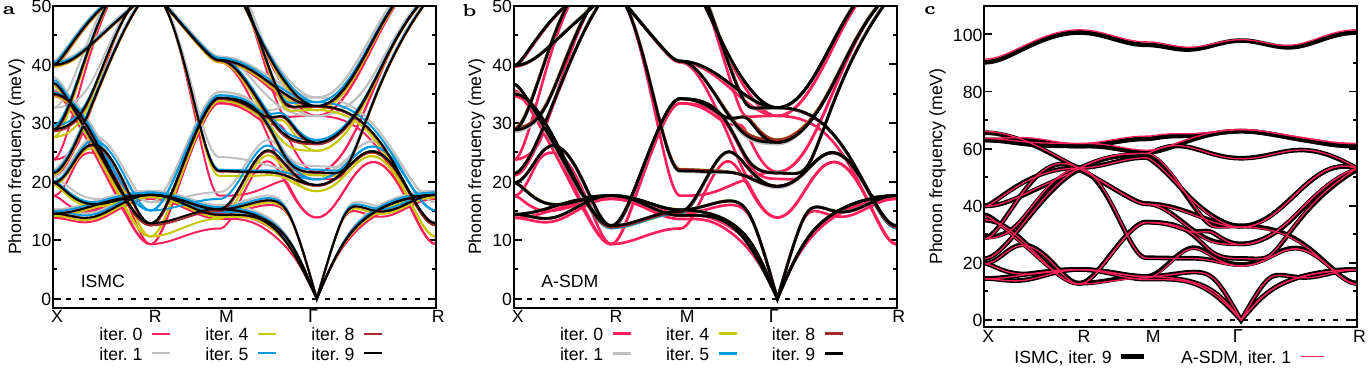}
\caption{ Iterative convergence of the phonon dispersion of cubic SrTiO$_3$ at 700~K calculated by (a) 
ISMC without mixing and (b) the A-SDM with linear mixing. 
In both cases, iteration 0 represents the symmetrized phonon dispersion at 0~K calculated for 
the polymorphous structure.
(c) Self-consistent phonon dispersion 
of SrTiO$_3$ as obtained at iteration 9 by ISMC (black) and 
iteration 1 by the A-SDM (red). 
All calculations refer to 2$\times$2$\times$2 supercells.
\label{fig10}}
\end{figure*}

If step 1 in the main procedure is skipped, the initial set of IFCs, $C_{p \k \a, p' \k' \a'}$, 
might be ill-defined, resulting to imaginary phonon frequencies. This might happen if the static equilibrium 
positions of the nuclei do not define the global minimum of the potential [e.g. green curve in Fig.~\ref{fig2}(b)]. 
Accounting for the global minimum [e.g. red curve in Fig.~\ref{fig2}(b)] yields
dynamically stable phonons. In systems with a multi-well PES featuring local saddle points, one 
needs to release the atoms away from their high-symmetry positions, usually defining a local maximum, 
and perform a tight relaxation of atomic positions (see also Sec.~\ref{comp.details}). This, in turn, 
leads to a low-symmetry distorted configuration similar to the strategy used in Ref.~[\onlinecite{Zhao_Zunger2020}] 
to explore polymorphism in metal halide and oxide perovskites. 
{\it We propose that the phonons calculated for the ground state polymorphous structure constitute 
the best starting point in the A-SDM. }
In general, the starting guess of the A-SDM self-consistent procedure
 is not necessarily unique, and one could resort to other choices (e.g. setting an electronic
temperature high enough) that yield positive definite IFCs; however we consider the choice of the 
polymorphous structure as the most physically relevant one. Furthermore, if the harmonic approximation 
gives an initial positive definite IFCs matrix (see for example Ref.~[\onlinecite{arxiv.2212.09789}]), 
then the iterative procedure can start without exploring the polymorphous network.

Starting the iterative procedure from ${\bf C}_0$ computed for the 
polymorphous network guarantees dynamically stable phonons. However, if iterative mixing 
in step 7 is not followed, then 
${\bf C}_j(T)$ might be ill-defined at later iterations, giving phonon instabilities. 
To deal with this problem one can set the imaginary frequencies to real and prepare the configuration for the
next iteration, as usually performed in stochastic implementations of the SCP theory~\cite{Mingo_2021}. 
However, this practise might deteriorate the stability of the whole approach.
Another issue is that the solution of ${\bf C}_j(T)$ might exhibit 
an oscillatory behavior, slowing down, or even degrading, self-consistency. 
In this case one can perform mixing of the oscillatory solutions 
of ${\bf C}_j(T)$ and ${\bf C}_{j-1}(T)$ and continue the iterative procedure.
We encountered this issue in our preliminary tests using the Monte Carlo 
approach discussed below.

The oscillatory behavior and unstable solutions can be avoided, a priori, if we employ
an iterative mixing of the IFCs (step 7 of the main procedure). 
Importantly, this approach is very beneficial for reducing statistical 
errors as well as for accelerating the convergence rate considerably. To demonstrate this 
point, in Fig.~\ref{fig10} we compare the convergence performances of the 
importance sampling Monte Carlo (ISMC)~\cite{Zacharias2015} approach without considering mixing 
with the A-SDM using linear mixing for $\beta =0.5$. 
We take as an example the phonon dispersion of SrTiO$_3$ at $T= 700$~K. The ISMC framework refers 
to obtaining convergence of the IFCs with (i) the number of thermal 
configurations generated stochastically at each iteration [Eq.~\eqref{eq_SDM_TA}] and (ii) 
the number of iterations. We found that 9 iterations (10 configurations each) are enough 
for obtaining convergence, as shown in Fig.~\ref{fig10}(a). This amounts to a total of 90 calculations. 
We note that in iterations 4 and 5 of ISMC the system was trapped between two oscillatory solutions. 
In order to proceed to the next iteration we performed mixing of the corresponding IFCs.
The A-SDM framework refers to using only one ZG configuration at each iteration which is generated by 
IFCs obtained through step 7 of the main procedure. The convergence performance of
the A-SDM is shown in Fig.~\ref{fig10}(b). Apart from requiring a {\it single configuration} at each iteration, 
the A-SDM outperforms ISMC to the point that only a {\it signle iteration} is enough for self-consistency. 
Figure~\ref{fig10}(c) shows that the A-SDM and ISMC give almost identical dispersions 
(and free energies) at their 1$^{\rm st}$ and 9$^{\rm th}$ iterations, respectively.

Let us note that linear mixing is the simplest approach possible to introduce mixing between IFCs of different 
iterations. Its success for the systems studied here is remarkable, enabling convergence, essentially, with two sets of 
calculations. However, for more challenging or complex cases, other sophisticated schemes based 
on quasi-Newton methods, like the Broyden or Broyden–Fletcher–Goldfarb–Shanno (BFGS) algorithms~\cite{Broyden1965,Fletcher1987}, 
could be proved useful.

\subsection{Computational details} \label{comp.details}
All electronic structure calculations were performed within density functional theory (DFT) 
using plane waves basis sets as implemented in {\tt Quantum Espresso} (QE)~\cite{QE,QE_2}.
We employed the Perdew-Burke-Ernzerhof exchange-correlation functional
revised for solids (PBEsol)~\cite{Perdew_2008} and optimized norm-conserving Vanderbilt pseudopotentials from
the PseudoDojo library~\cite{Haman_2013,vanSetten2018}.
For cubic SrTiO$_3$, CsPbBr$_3$, CsPbI$_3$, and CsSnI$_3$ (space group $Pm\bar{3}m$) we set the kinetic energy 
cutoff to 120~Ry and sampled the Brillouin zone of the unit cells (5 atoms) and 2$\times$2$\times$2 supercells
with 6$\times$6$\times$6 and 3$\times$3$\times$3 uniform $\bk$-grids.
The calculations for the tetragonal phase of SrTiO$_3$ (space group $I4/mcm$) were performed 
with the same cutoff using 4$\times$4$\times$4 and 2$\times$2$\times$2 uniform $\bk$-grids for the unit cell 
(10 atoms) and 2$\times$2$\times$2 supercells, respectively.
For bcc Zr (space group $Im\bar{3}m$) we used a cutoff of 80 Ry and sampled the 
Brillouin zone of the unit cell, 3$\times$3$\times$3, 4$\times$4$\times$4, and 5$\times$5$\times$5 
supercells with 10$\times$10$\times$10, 4$\times$4$\times$4, 3$\times$3$\times$3, 
and 2$\times$2$\times$2 uniform $\bk$-grids.
The calculations of the high-symmetry (monomorphous) structures were performed using the unit cells 
with the nuclei clamped at their Wyckoff positions.
The relaxed lattice constants of cubic SrTiO$_3$, CsPbBr$_3$, CsPbI$_3$, CsSnI$_3$, and Zr are found to be
3.889, 5.874, 6.251, 6.141, and 3.517~\AA, respectively; for tetragonal SrTiO$_3$ our calculations
yield $a=5.485$ and $b=7.806$~\AA.
These parameters were used for all calculations, unless
specified otherwise. For the orthorhombic CsPbBr$_3$ (space group $Pnma$), we employed the unit cell containing 20 atoms and
a 3$\times$3$\times$3 uniform $\bk$-grid to find relaxed lattice constants of $a=7.971$, $b=8.397$, and $c=11.640$~\AA. 
The phonons were computed using a 2$\times$2$\times$2 supercell and a 2$\times$2$\times$2 uniform $\bk$-grid. 

To obtain DFT ground state geometries of the cubic and tetragonal phases
(the polymorphous networks), we started from the monomorphous network
and displaced the nuclei away from their Wyckoff positions using ZG displacements 
[Eq.~\eqref{eq.realdtau_method00}] at $T=0$~K and the harmonic phonons.
Then, we allowed the system to relax until the residual force component per atom was 
less than 3$\times$10$^{-4}$ eV/\AA. We found that displacing the atoms along the soft modes only by switching 
their phonon frequencies to the real axis is a much more efficient strategy to obtain the polymorphous geometry. 
We made this choice because (i) ZG displacements generated for the monomorphous structure are otherwise not defined 
for the soft modes and (ii) the system is brought closer to its ground state. We stress that we tested various
initial sets of displacements (random nudges and ZG displacements by changing signs or temperatures) that led to 
different polymorphous networks; however, all yield the same ground state energy and phonons~\cite{Zacharias2023}.


The IFCs of the monomorphous and polymorphous supercell configurations were evaluated 
using the FPM~\cite{Kunc_Martin,phonopy}. 
The phonon eigenmodes and eigenfrequencies at each $\bq$-point  
were obtained by means of Fourier interpolation of the corresponding dynamical matrices. 
Corrections on the phonon dispersion due to long-range dipole-dipole interactions 
were included via the linear response approach described in Ref.~[\onlinecite{Gonze1997}].
High frequency dielectric constants and atomic Born effective charges were calculated using 
DFPT~\cite{Baroni2001} as implemented in QE.
ZG displacements in 2$\times$2$\times$2 supercells were employed for calculating temperature-dependent anharmonic phonons. 
These displacements were generated via SDM~\cite{Zacharias_2016,Zacharias_2020} as implemented 
in the {\tt EPW/ZG} code~\cite{Hyungjun2023}. For their construction we (i) used the phonons 
of a zone-centered $\bq$-grid commensurate with the desired supercell size, (ii) took into account dipole-dipole interaction 
corrections, and (iii) apply a Berry connection between the eigenmodes. 

The {\tt ZG.x} code implementing the A-SDM procedure with the FPM is available at the  
{\tt EPW/ZG} module~\cite{gitlab_ZG}.
A-SDM phonon dispersions were obtained using 2$\times$2$\times$2 supercells for the perovskite structures
and 3$\times$3$\times$3/4$\times$4$\times$4/5$\times$5$\times$5 supercells for bcc Zr. 
At each iteration, including iteration 0 for the polymorphous structure, we enforced 
the crystal's symmetry operations on the IFCs.
Setting the mixing parameter $\beta$ to 0.5, we found that a couple of iterations is enough 
to obtain reasonable convergence and a maximum of 3-4 iterations to achieve full convergence. 
To ensure high accuracy for the temperature dependence of the TO1 and TO2 modes of SrTiO$_3$, 
shown in Fig.~\ref{fig5}(e), we employed 10 iterations as default.
Long-range corrections in A-SDM phonon dispersions and ZG displacements were accounted for using 
the Born effective charges and dielectric constants of the polymorphous structures (see Table~\ref{table.3}). 


\section{Conclusions and outlook} \label{conclusions}

In this paper, we have developed the A-SDM for the efficient treatment of anharmonicity. 
Similarly to previous nonperturbative methodologies, the A-SDM is based on the self-consistent phonon theory.
The key feature introduced here is that special displacements can completely replace the cumbersome stochastic
sampling of thermally displaced configurations, showing that a single ZG configuration is sufficient
to compute temperature-dependent phonon dispersions at each iteration. Furthermore, we have demonstrated that the 
stability and efficiency of the self-consistent phonon approach can be considerably improved by means of an iterative 
linear mixing scheme. 
Another essential ingredient for accelerating convergence in the A-SDM is the initialization
of the iterative procedure with interatomic force constants calculated for the polymorphous ground state network.
Considered all together, we are suggesting, essentially, that only a couple of ZG configurations are needed to 
describe anharmonicity; one to obtain the polymorphous structure and one to obtain the phonon dispersion at a 
given temperature. 

We have benchmarked our methodology for the perovskites SrTiO$_3$, CsPbBr$_3$, CsPbI$_3$, 
and CsSnI$_3$, which exhibit strongly anharmonic multi-well potential energy surfaces. 
From the calculated phonon dispersions, we have extracted temperature-dependent frequencies of soft modes 
at high-symmetry points and obtain excellent agreement with previous first-principles studies.
We have also evaluated the phonon dispersion of the high temperature bcc phase of Zr, yielding 
good agreement with experiments and previous calculations. Our findings suggest that the phonons 
calculated for polymorphous SrTiO$_3$ and Zr at 0~K provide a good approximation to explore 
anharmonicity in these systems. We emphasize that evaluating the phonons of polymorphous networks relies 
on standard techniques that are routinely used for harmonic systems. 

This work is also considered as the upgrade of SDM to anharmonic materials. SDM has been originally developed 
to automatically incorporate the effect of electron-phonon coupling in supercell calculations of 
band structures and optical spectra~\cite{Zacharias_2016}, and extended to other properties like 
transport coefficients~\cite{Gunst2017}. The present study address the constraint 
of the harmonic approximation and opens the way for a unified treatment to electron-phonon coupling and anharmonicity
in systems exhibiting a complex PES. 
In fact, here, we have demonstrated that the A-SDM is very effective for computing  
well-defined phonon dispersions of technologically important materials
like the metal halide perovskites CsPbBr$_3$, CsPbI$_3$, and CsSnI$_3$. Anharmonicity is ubiquitous 
in this class of materials~\cite{Marronnier2017,Marronnier2018,Katan2018,Ferreira2020} and 
interfacing efficient computations of phonons and electron-phonon couplings can shed light 
on their unexplored equilibrium or nonequilibrium properties~\cite{Zhang2023}. 
Furthermore, we stress that the feature of computing thermodynamically stable phonon spectra via the A-SDM can
be straightforwardly exploited by state-of-the-art perturbative electron-phonon calculations~\cite{Hyungjun2023}.

Finally, it should be possible to extend the A-SDM for the computation of phonon transport properties dominated
by phonon–phonon interactions. For example, calculations of phonon relaxation times and thermal conductivity via the 
Green-Kubo approach should be within reach~\cite{Carbogno_2017}.
Furthermore, we expect the A-SDM to find applications in the investigation of ultrafast electron and phonon dynamics 
obtained by the solution of the coupled time-dependent Boltzmann equations~\cite{Caruso2021,Britt2022}. 
Given the generality and efficiency of our methodology, we also expect to be perfectly suited for 
high-throughput calculations of several electron-phonon and anharmonic properties of materials at finite temperatures. 

Calculations that lead to the results of this
study are available on the NOMAD repository~\cite{nomad_doi}.

\acknowledgments

M.Z. acknowledges funding from the European Union’s Horizon 2020 research and innovation programme
under the Marie Skłodowska-Curie Grant Agreement No. 899546. 
This research was also funded by the European Union (project ULTRA-2DPK / HORIZON-MSCA-2022-PF-01 /
Grant Agreement No. 101106654). Views and opinions expressed are however those of the authors only and do not necessarily 
reflect those of the European Union or the European Commission. Neither the European Union nor the 
granting authority can be held responsible for them.
J.E. acknowledges financial support from the Institut Universitaire de France.
F.G. was supported by the National Science Foundation under CSSI Grant No. 2103991 and DMREF Grant No. 2119555.
We acknowledge that the results of this research have been achieved using the DECI resource 
Prometheus at CYFRONET in Poland [https://www.cyfronet.pl/] with support from the PRACE aisbl
and HPC resources from the Texas Advanced Computing Center (TACC) at The University of Texas at Austin
[http://www.tacc.utexas.edu].

\appendix 

\section{Proof of Eq.~\eqref{eq.IFC_T}} \label{app.proof}
In this Appendix we derive Eq.~\eqref{eq.IFC_T} following the work in Ref.~[\onlinecite{Bianco2017}].

The expression in Eq.~\eqref{eq.FE} is considered as the trial free energy which, according to the Gibbs-Bogolyubov 
inequality, provides an upper limit of the system's free energy.
Hence, the derivation of Eq.~\eqref{eq.IFC_T} relies on the minimization of the free energy with respect to 
the matrix of IFCs and, thus, finding the solution:
\begin{eqnarray}  \label{eq.bla}
\frac{\partial F (T) }{ \partial {C}_{\k\a,\k' \a'}} = 0,
\end{eqnarray}
where for the sake of notation clarity we use a $\Gamma$-point formalism and drop the unit cell index $p$.

We consider first the derivative of $\braket{U}_T$ and use the chain rule with respect to $\tau_{\k \a}$ to obtain: 
\begin{eqnarray}\label{eqd_2}
\frac{\partial \braket{{U}}_T}{\partial C_{ \k \a,  \k' \a'}} = 
  \prod_{\nu} \!\int\! \frac{dz_{\nu} }{u_{\nu} \sqrt{2\pi}  } 
    e^{-\frac{z_{ \nu}^2}{2 u^2_{ \nu}}} 
    \sum_{ k'' \a''} \frac{\partial U} {\partial \tau_{ \k'' \a''}} 
    \frac{\partial \tau_{ \k'' \a''}} {\partial C_{\k \a, \k' \a'}}, \nonumber \\
\end{eqnarray}
where we express the thermal average in its multivariate Gaussian integral form appearing in Eq.~\eqref{eq_SDM_TA}.
We employ 
$\tau_{\k \a} = (M_0 / M_\k)^{1/2} \sum_\nu e_{\k \a,\nu} z_\nu $, representing the normal coordinates transformation, 
and apply the change of variables $z_\nu = \tilde{z}_\nu u_\nu$ to rewrite Eq.~\eqref{eqd_2} as: 
\begin{widetext}
\begin{eqnarray}\label{eqd_3b}
\frac{\partial \braket{{U}}_T}{\partial C_{ \k \a, \k' \a'}} = 
  \sum_{ k'' \a''} \bigg(\frac{M_0}{M_{\k''}}\bigg)^{1/2}  \sum_\mu  
\frac{\partial \big[e_{\k'' \a'',\mu} u_\mu \big]} {\partial C_{\k \a, \k' \a'}}
\prod_{\nu \neq \mu} 
 \!\int\! \frac{d\tilde{z}_{\nu} }{ \sqrt{2\pi}  } \, 
    e^{-\frac{\tilde{z}_{ \nu}^2}{2 }} 
   \! \int \! \frac{d\tilde{z}_{\mu} }{\sqrt{2\pi}  }  \,
    e^{-\frac{\tilde{z}_{ \mu}^2}{2 }} \tilde{z}_\mu
    \frac{\partial U} {\partial \tau_{ \k'' \a''}}. 
\end{eqnarray}
We perform integration by parts with respect to $\tilde{z}_\mu$, take
the limit $\lim_{\rightarrow \infty} e^{-z^2} = 0$, and use $\partial / \partial \tilde{z}_\mu 
=  \sum_{\k \a} (M_0/M_\k)^{1/2} e_{\k \a,\mu} u_\mu \, \partial / \partial \tau_{\k \a}$
to obtain the following result after some straightforward algebra:
\begin{eqnarray} \label{eqd_finalish}
\frac{\partial \braket{{U}}_T}{\partial C_{ \k \a, \k' \a'}} &=&
 \frac{1}{2} \sum_{\substack{\k'' \a'' \\ \k''' \a'''}} \frac{M_0}{ \sqrt{M_{\k''} M_{\k'''}}} \sum_\mu 
\frac{\partial \big[e_{\k'' \a'',\mu} e_{\k''' \a''',\mu}\,u^2_\mu \big]} {\partial C_{\k \a, \k' \a'}} 
   \Braket{ \frac{\partial U} {\partial \tau_{ \k'' \a''} \partial \tau_{ \k''' \a'''}}}_T.
\end{eqnarray}
\end{widetext}

Now we consider the derivative of the harmonic vibrational free energy $F_{\rm vib}(T)$
[term appearing in the second line of Eq.~\eqref{eq.FE}] which we rewrite here for 
convenience: 
\begin{eqnarray} \label{eq.FEvib}
F_{\rm vib}(T) = \sum_{ \nu} \bigg[ \frac{\hbar \omega_{\nu}}{2} - k_{\rm B} T\, {\rm ln} (1+ n_\nu) \bigg].
\end{eqnarray}
Employing the chain rule with respect to $\w^2_\mu$ gives: 
\begin{eqnarray} \label{eq_milestone_1}
  \frac{\partial{F_{\rm vib}(T)}} {\partial C_{ \k \a, \k' \a'} } &=& 
\sum_\mu \frac{M_0}{2} u^2_\mu \frac{\partial \w^2_\mu} {\partial C_{\k \a, \k' \a'} },
\end{eqnarray}
where we have used 
$\partial n_{\nu} / \partial \w_\nu^2 =- \hbar/ (2 k_{\rm B} T \w_\nu) \, n_{\nu} (n_{\nu} + 1)$.
By combining Eqs.~\eqref{eq.Dynmat} and~\eqref{eq.eig_dyn_mat}, and using the orthonormality relations 
of the phonon eigenvectors~\cite{Giustino_2017} we have:
\begin{eqnarray} \label{eq_w2}
  \w^2_\mu = \sum_{\k \a, \k' \a'} \frac{C_{\k \a, \k' \a'}}{\sqrt{M_\k M_{\k'}}} e_{\k \a,\mu} \, e_{\k' \a',\mu}, 
\end{eqnarray}
that allows us to express Eq.~\eqref{eq_milestone_1} as: 
\begin{eqnarray} \label{eq_milestone_2}
   \frac{\partial{F_{\rm vib}(T)}} {\partial C_{ \k \a, \k' \a'} } =
 \frac{1}{2} \frac{M_0}{\sqrt{M_\k M_{\k'}}} \sum_{\mu} e_{\k \a,\mu}\, e_{\k' \a',\mu} u^2_\mu. 
\end{eqnarray}

Now we consider the vibrational energy of the effective harmonic Hamiltonian $U_{\rm h}(T)$
which we also rewrite here for convenience:
\begin{eqnarray} \label{eq.Ueff}
U_{\rm h}(T) = \frac{M_0}{2} \sum_{\mu} \omega^2_{\mu} u^2_{ \mu}. 
\end{eqnarray} 
By combining Eqs.~\eqref{eq.Ueff} and~\eqref{eq_w2}, and
taking the derivative with respect to ${C}_{\k\a, \k' \a'}$ yields the following result: 
\begin{widetext}
\begin{eqnarray} \label{eq.milestone_3}
 \frac{\partial U_{\rm h}(T)} {\partial C_{ \k \a, \k' \a'}}& = &
  \frac{1}{2}  \frac{M_0}{\sqrt{M_\k M_{\k'}}} \sum_{\mu} e_{\k \a,\mu} e_{\k' \a',\mu} u^2_\mu + 
 \sum_{\substack{\k'' \a'' \\ \k''' \a'''}} \frac{C_{ \k'' \a'', \k''' \a'''}}{2}  \frac{M_0}{\sqrt{M_{\k''} M_{\k'''}}}
 \sum_{\mu}   \frac{\partial \big[ e_{\k'' \a'',\mu} e_{\k''' \a''',\mu} u^2_\mu \big] } {\partial C_{ \k \a, \k' \a'} }
\end{eqnarray}

Finally, we combine Eqs.~\eqref{eq.FE}, \eqref{eq.Ueff}, and \eqref{eq.FEvib} together 
with the expressions in Eqs.~\eqref{eqd_finalish}, \eqref{eq.milestone_3}, and~\eqref{eq_milestone_2} to obtain:
\begin{eqnarray} \label{eq.bla2}
\frac{\partial F(T)}{\partial {C}_{ \k \a, \k' \a'}} &=& 
\frac{\partial \braket{U}_T}{\partial C_{ \k \a, \k' \a'}} - \frac{\partial U_{\rm h}(T)}{\partial C_{ \k \a, \k' \a'}}
+ \frac{\partial{F_{\rm vib}(T)}} {\partial C_{ \k \a, \k' \a'} }  
\nonumber
\\ &=& \frac{1}{2} \sum_{\substack{\k'' \a'' \\ \k''' \a'''}} \frac{M_0}{ \sqrt{M_{\k''} M_{\k'''}}} \sum_\mu 
\frac{\partial \big[e_{\k'' \a'',\mu} e_{\k''' \a''',\mu}\,u^2_{ \mu} \big]}{\partial C_{\k \a, \k' \a'}}
\bigg[ \Braket{ \frac{\partial U} {\partial \tau_{ \k'' \a''} \partial \tau_{ \k''' \a'''}}}_T - C_{ \k'' \a'', \k''' \a'''} \bigg].
\end{eqnarray}
\end{widetext}
Hence, in order to satisfy Eq.~\eqref{eq.bla}, the term inside the square brackets of Eq.~\eqref{eq.bla2}
must be zero. This, essentially, completes the proof of Eq.~\eqref{eq.IFC_T} which requires evaluating 
the matrix of IFCs at each temperature in a self-consistent fashion.

\bibliography{references}{}

\end{document}